%% file: main.tex
\documentclass[a4paper,11pt]{article}

\pdfoutput=1
\usepackage{jheppub} 
\usepackage{subcaption}
\usepackage{amsmath}
\usepackage{lineno}
\usepackage{amsmath}
\usepackage{comment}
\usepackage[usenames,dvipsnames,table]{xcolor}

\usepackage{indentfirst}
\usepackage{xspace}

\usepackage{booktabs}
\usepackage{siunitx}
\usepackage{cancel}

\usepackage{amssymb}
\usepackage{amsmath}
\usepackage{subcaption}

\newcommand{\orca}{KM3NeT/ORCA}
\newcommand{\osix}{ORCA6}

\newcommand{\nutau}{$\nu_{\tau}$}

\newcommand{\tauCC}{$\nu_\tau$~CC }

\newcommand{\bfpm}[3]{$#1^{+#2}_{-#3}$ }
\newcommand{\threeflav}{3$\nu$-paradigm}
\newcommand{\upmns}{$U_{\textrm{PMNS}}$}
\newcommand{\deltacp}{$\delta_{\textrm{CP}}$}
\newcommand{\Stau}{\emph{S}$_{\tau}$}
\newcommand{\Vnc}{$V_{\textrm{NC}}$}

\abstract{Oscillations of atmospheric muon and electron neutrinos produce tau neutrinos with energies in the GeV range, which  can be observed by the ORCA detector of the KM3NeT neutrino telescope in the Mediterranean Sea. First measurements with ORCA6, an early subarray corresponding to about 5$\%$ of the final detector, are presented.
A sample of 5828 neutrino candidates has been selected from the analysed exposure of 433 kton-years.
The $\nu_\tau$~normalisation, defined as the ratio between the number of observed and expected tau neutrino events, is measured to be $S_\tau = 0.48^{+0.5}_{-0.33}$. This translates into a \nutau~charged-current cross section measurement of $\sigma_{\tau}^{\text{meas}} = (2.5 ^{+2.6}_{-1.8}) \times 10^{-38}$ cm$^{2}$ nucleon$^{-1}$ at the median $\nu_\tau$ energy of 20.3\,GeV. The result is consistent with the measurements of other experiments.
In addition, the current limit on the non-unitarity parameter affecting the $\tau$-row of the neutrino mixing matrix was improved, with $\alpha_{33}>$ 0.95 at the 95$\%$ confidence level.
}

\begin{document}
\title{Study of tau neutrinos and non-unitary neutrino mixing with the first six detection units of KM3NeT/ORCA}

\include{tex/authors}

\maketitle

\flushbottom
\raggedbottom

\input{tex/introduction}
\input{tex/orca}
\input{tex/physics}
\input{tex/selection}
\input{tex/analysis}
\input{tex/results}
\input{tex/conclusions}
\input{tex/acknowledgements}

\bibliographystyle{JHEP}
\bibliography{main.bib}

\end{document}

%% file: tex/authors.tex
\newcommand\blfootnote[1]{%
  \begingroup
  \renewcommand\thefootnote{}\footnote{#1}%
  \addtocounter{footnote}{-1}%
  \endgroup
}

\newcounter{daggerfootnote}
\newcommand*{\daggerfootnote}[1]{%
    \setcounter{daggerfootnote}{\value{footnote}}%
    \renewcommand*{\thefootnote}{\fnsymbol{footnote}}%
    \footnote[2]{#1}%
    \setcounter{footnote}{\value{daggerfootnote}}%
    \renewcommand*{\thefootnote}{\arabic{footnote}}%
}

\author[a]{S.~Aiello, }
\author[b,be]{A.~Albert, }
\author[c]{A.\,R.~Alhebsi, }
\author[d]{M.~Alshamsi, }
\author[e]{S. Alves Garre, }
\author[g,f]{A. Ambrosone, }
\author[h]{F.~Ameli, }
\author[i]{M.~Andre, }
\author[j]{L.~Aphecetche, }
\author[k]{M. Ardid
% ...
}
% ...
\emailAdd{km3net-pc@km3net.de}
\affiliation[a]{INFN, Sezione di Catania, (INFN-CT) Via Santa Sofia 64, Catania, 95123 Italy}
\affiliation[b]{Universit{\'e}~de~Strasbourg,~CNRS,~IPHC~UMR~7178,~F-67000~Strasbourg,~France}
\affiliation[c]{Khalifa University of Science and Technology, Department of Physics, PO Box 127788, Abu Dhabi,   United Arab Emirates}
\affiliation[d]{Aix~Marseille~Univ,~CNRS/IN2P3,~CPPM,~Marseille,~France}
\affiliation[e]{IFIC - Instituto de F{\'\i}sica Corpuscular (CSIC - Universitat de Val{\`e}ncia), c/Catedr{\'a}tico Jos{\'e} Beltr{\'a}n, 2, 46980 Paterna, Valencia, Spain}
\affiliation[f]{INFN, Sezione di Napoli, Complesso Universitario di Monte S. Angelo, Via Cintia ed. G, Napoli, 80126 Italy}
\affiliation[g]{Universit{\`a} di Napoli ``Federico II'', Dip. Scienze Fisiche ``E. Pancini'', Complesso Universitario di Monte S. Angelo, Via Cintia ed. G, Napoli, 80126 Italy}
\affiliation[h]{INFN, Sezione di Roma, Piazzale Aldo Moro 2, Roma, 00185 Italy}
\affiliation[i]{Universitat Polit{\`e}cnica de Catalunya, Laboratori d'Aplicacions Bioac{\'u}stiques, Centre Tecnol{\`o}gic de Vilanova i la Geltr{\'u}, Avda. Rambla Exposici{\'o}, s/n, Vilanova i la Geltr{\'u}, 08800 Spain}
\affiliation[j]{Subatech, IMT Atlantique, IN2P3-CNRS, Nantes Universit{\'e}, 4 rue Alfred Kastler - La Chantrerie, Nantes, BP 20722 44307 France}
\affiliation[k]{Universitat Polit{\`e}cnica de Val{\`e}ncia, Instituto de Investigaci{\'o}n para la Gesti{\'o}n Integrada de las Zonas Costeras, C/ Paranimf, 1, Gandia, 46730 Spain}
% ...
\affiliation[be]{Universit{\'e} de Haute Alsace, rue des Fr{\`e}res Lumi{\`e}re, 68093 Mulhouse Cedex, France}
% ...

%% file: tex/introduction.tex
\section{Introduction}
\label{sec:intro}

The experimental observation of neutrino oscillations in 1998~\cite{SK0} provided evidence that neutrinos have non-zero mass. Neutrino flavour eigenstates ($\nu_e$, $\nu_{\mu}$, $\nu_{\tau}$) are linear combinations of the mass eigenstates ($\nu_1$, $\nu_2$, $\nu_3$) expressed through the Pontecorvo-Maki-Nakagawa-Sakata (PMNS) mixing matrix~\cite{Pontecorvo:1957cp, Maki:1962mu}. The oscillation probabilities are a function of the distance travelled by neutrinos, their energy, the mixing matrix and the mass-squared differences between the mass states, $\Delta m^2_{ij} =m^2_{i} -m^2_{j}$. However, the Standard Model of elementary particles does not provide a mechanism describing the origin of such masses, and neutrino physics is a first field showing a clear necessity for physics beyond the Standard Model.

From the experimental data gathered so far, a coherent picture of the neutrino mixing has been achieved. 
The PMNS matrix (\upmns) is defined by three mixing angles ($\theta_{12}$, $\theta_{13}$, $\theta_{23}$), a charge-parity (CP) violation phase (\deltacp), and is assumed to be unitary. 
The values of the oscillation parameters are currently known with percentage-level precision~\cite{nufit5_0, Capozzi2003, deSalas:2020pgw}. Critical questions regarding the oscillation mechanism still need to be addressed. Providing those answers, with the determination of the neutrino mass ordering by clarifying the sign of $\Delta m^2_{31}$ and the precise measurement of the $\delta_{\textrm{CP}}$-phase to investigate the presence of CP violation in the leptonic sector, would open a way to explore new physics. While the quark mixing matrix has been extensively tested with robust precision, the unitarity of the PMNS matrix, assumed in many oscillation analyses~\cite{NOvA:2023iam, T2K2023}, still has to be experimentally proven.
Under the PMNS matrix unitarity hypothesis, solar~\cite{SNO1, SNO2}, reactor~\cite{DB1, DB2, DoubleChooz, RENO}, and atmospheric~\cite{SK0, SK1, IC-DeepC} neutrino experiments mostly exploited the $\nu_e$, $\bar{\nu}_e$, and $\nu_{\mu}$, $\bar{\nu}_{\mu}$ disappearance channels, whereas the high-statistics in $\nu_{e}$ and $\bar{\nu}_{e}$ appearance channels have been investigated in long-baseline accelerator experiments~\cite{NOvA:2023iam, T2K2023}. 

The detection and study of $\tau$\xspace neutrinos\footnote{In this paper, the term \emph{neutrino} generally refers to both neutrinos and antineutrinos if not specified otherwise.} is challenging. Their existence was suggested in 1977~\cite{nutau_pred}, while their first experimental observation was reported only in 2000~\cite{DONuT2000}. 
The main detection channel is the $\tau$ lepton produced via charged-current (CC) interactions, with a production energy threshold of about 3.5\,GeV and a suppressed cross section near this threshold. At accelerators, \nutau~are created through the decays of heavy charmed hadrons. In the atmosphere, \nutau~can be observed mainly via the oscillation channel, largely taking place above the production threshold, which becomes accessible for a baseline longer than $\sim$2000\,km. However, the short lifetime of the $\tau$-lepton requires excellent particle identification techniques for a clean \nutau~event reconstruction~\cite{DONuT2000, opera}. Overall, the main current limitation is the statistics. 
Therefore, in contrast with the current precision on the PMNS matrix elements in the electron and muon rows, the $\tau$-row, remains the least constrained. This difference is enhanced when allowing for non-unitarity~\cite{Parke_2016, Zhuojun, Denton:2021}.
In the standard three neutrino framework (\threeflav), 12 conditions on the PMNS matrix elements exist; six of them are given by the sum of the elements squared in each row (referred to as \emph{normalisation} and expected to be equal to 1). 
A plethora of beyond the Standard Model extensions are proposed to validate the \threeflav\xspace. Moreover, the tau neutrino interaction model and cross section currently lack precise experimental data, in order to validate the theoretical prediction~\cite{SK_tau}. In this context, the \nutau~appearance, as direct detection of \nutau\xspace from the oscillation mechanism, is a golden channel to constrain the \nutau~CC cross section.

Several running and next-generation neutrino experiments are exploiting different techniques to reach a deeper understanding of tau neutrinos~\cite{MammenAbraham:2022xoc} from the GeV to the EeV energy scale. 
This paper investigates the properties of atmospheric tau neutrinos, using data collected with a partial configuration of the KM3NeT/ORCA neutrino telescope (5$\%$ of the nominal volume, in the following referred to as ORCA6), from January 2020 to November 2021. 
Currently under construction in the Mediterranean Sea offshore the south of France, KM3NeT/ORCA is optimised for the study atmospheric neutrino oscillations in the 1--100\,GeV energy range. 
Considering the foreseen final instrumented volume, the strong matter effect, the accessible low-energy detection threshold, and the good angular and energy resolutions, the $\nu_{\mu} \rightarrow \nu_{e}$ appearance channel will allow for probing the neutrino mass ordering by studying the resonant oscillation in the Earth's mantle~\cite{nmo2022}. The $\nu_{\mu}$ disappearance is the dominant channel and contributes to a precise measurement of the atmospheric oscillation parameters $\Delta m^2_{31}$ and $\theta_{23}$. Additionally, KM3NeT/ORCA will have good sensitivity to the \nutau~appearance channel thanks to its large accessible statistics, $\sim$3000~\nutau~CC/year in the final detector volume~\cite{LoI2016, nmo2022}. 

The structure of the paper is organised as follows. The KM3NeT/ORCA telescope and the used data sample are described in section~\ref{sec:orca_exp}; Monte Carlo simulations and the data taking conditions are described in section~\ref{sec:data_taking} and~\ref{sec:simu}. In section~\ref{sec:physics}, the theoretical aspects of the \nutau~appearance channel investigation are explained, with a focus on the impact on the \nutau\xspace cross section measurement in section~\ref{sec:cc_ccnc}  and on the theory behind the non-unitarity test of the neutrino mixing matrix in section~\ref{sec:NUNM}. The event selection and analysis method are reported in section~\ref{sec:ev} and section~\ref{sec:ana}, respectively. The results are summarised in section~\ref{sec:res}: the cross section measurement and results on non-unitarity constraints are reported in section~\ref{sec:cc} and~\ref{sec:nc}, respectively. A conclusive summary describing the physics contribution of ORCA6 analysis to already existing results is reported in section~\ref{sec:conc}.

%% file: tex/orca.tex
\section{The KM3NeT/ORCA detector}
\label{sec:orca_exp}

The KM3NeT Collaboration is building two large-volume water Cherenkov neutrino detectors in the Mediterranean Sea: KM3NeT/ARCA (Astroparticle Research with Cosmics in the Abyss) and KM3NeT/ORCA (Oscillation Research with Cosmics in the Abyss). The two detectors have complementary physics programs and explore neutrinos in a wide energy range, from MeV to PeV~\cite{LoI2016}. KM3NeT/ARCA, under construction 100 km offshore Portopalo di Capo Passero (Sicily, Italy), at a depth of 3500 m, is optimised for high energy astrophysical neutrino detection in the TeV-PeV energy range. KM3NeT/ORCA is being built at 2450 m depth, 40\,km offshore Toulon (France). Both detectors consist of a 3-dimensional array of photomultiplier tubes (PMTs) collecting the Cherenkov radiation induced in water by relativistic charged particles emerging from neutrino interactions with the seawater and seabed. The detector's key component is a spherical structure, the Digital Optical Module (DOM), hosting 31 PMTs, ensuring 4$\pi$ coverage for the detection of Cherenkov photons~\cite{DOMpaper}. A set of 18 DOMs are integrated into a Detection Unit (DU), kept vertically from the seabed thanks to buoys completing the top of each DU and the buoyancy of the DOMs themselves. The horizontal and vertical distances between optical sensors in the two neutrino telescopes are optimised to facilitate their main physics goals. 
In order to detect GeV neutrinos with KM3NeT/ORCA, its vertical inter-DOM and horizontal inter-DU distances are $\sim$9\,m and $\sim$20\,m, respectively. The number of installed DUs determines the fiducial detector mass and will reach about 7\,Mton (115 DUs) in the final design geometry. 
Thanks to its modular structure, \orca\xspace can detect neutrinos with a partially instrumented volume, which at the time of writing, covers about 20$\%$ of the nominal volume.

\subsection{Data taking, event reconstruction, and trigger algorithms}
\label{sec:data_taking}

Atmospheric neutrinos are the main signal in KM3NeT/ORCA. They are the decay products of kaons and pions created in the collision between cosmic rays and nuclei in the Earth's atmosphere. Contaminating the neutrino signature, two main background sources are present: atmospheric muons and environmental optical noise. The former consists of downgoing muons from extensive air showers. The latter is mostly caused by the radioactive decay of $^{40}$K in seawater and bioluminescence present in the detector surroundings.

The KM3NeT detector readout system~\cite{Chiarusi2023} is based on the \emph{all-data-to-shore} concept, so that all analogue signals from the PMTs exceeding a tunable threshold (typically 0.3 photoelectrons) are digitised and sent to shore for real-time processing. The data stream is processed using different levels of trigger conditions, including time and geometrical causality of the triggered hits. The stored optical data (the ``\textit{event}'') contains the time and the time-over-threshold of each analogue pulse (jointly referred to as ``\textit{hit}'') in a time window of a few $\mu s$ around those hits that satisfy one of the trigger conditions. This procedure provides a 10$^5$ reduction factor to the data stream before storing it on disk.

The event reconstruction is based on maximum-likelihood algorithms to extracting the relevant information of each event based on the space-time distribution of PMT signals in the detector. Depending on the neutrino flavour and the type of interactions (CC and neutral-current, NC), two main event topologies can be identified: \emph{track}-like and \emph{shower}-like. The first topology occurs every time a muon is produced as a secondary particle; the latter, occurs in all the other cases. The maximum likelihood algorithms assume, for each event, either a track or a shower hypothesis and provide an estimation of the direction and energy~\cite{ORCA6}. 
In ORCA6, 70$\%$ of track-like events are correctly reconstructed for events with energy above $\sim$30\,GeV; below this energy, events are identified as shower-like.

\subsection{Data processing and quality selection}
\label{sec:rbr}

The data stream is divided into \emph{runs}, with a typical duration of about 6 hours. The run duration is chosen to facilitate file transfer to KM3NeT Tier-1 computing centres and subsequent analysis. Summary data containing the rates of all PMTs in the detector are stored with a sampling frequency of 10 Hz. This information is used in the simulation and the reconstruction of the events on a run basis. 

For the ORCA6 dataset, additional data quality criteria are applied to exclude runs acquired in particularly unstable conditions, due to periods of high bioluminescence, timing accuracy issues, and high trigger rates  corresponding to 4.2\% of the raw data sample (more details on the quality criteria applied for the run selection are given in~\cite{ORCA6}).  
The results reported in this paper are obtained from a data sample including the reconstruction of both track-like and shower-like topologies, which corresponds to an exposure of 433\, kton-years (510 days of detector uptime).
Starting from this dataset, events are selected and classified to reject the main background and identify the neutrino sample as detailed in section~\ref{sec:ev}.

\subsection{Monte Carlo simulations}
\label{sec:simu}

Detailed Monte Carlo (MC) simulations of atmospheric neutrinos and atmospheric muons are used to evaluate the KM3NeT/ORCA detector response. 
A \emph{run-by-run} approach~\cite{rbr_antares} is used to mitigate the time-dependent impact of data taking. This approach exploits information on the active/inactive PMTs, the run setup, and the level of environmental optical background. 
The simulation chain~\cite{ORCA6} uses the GENIE-based~\cite{genie} software package gSeaGen~\cite{gSeaGen2020} for neutrino-induced interactions in seawater with the tune G18\_02a\_00\_000. Atmospheric neutrinos in the 1\,GeV--10\,TeV energy range are simulated; all neutrino flavours and both CC and NC interactions are considered. MC events are weighted according to the Honda model~\cite{honda}, using the atmospheric neutrino flux estimated in the Northern hemisphere (Fr\'ejus site). The path of neutrinos through the Earth is simulated using a density profile model based on the Preliminary Reference Earth Model (PREM)~\cite{prem}, consisting of 15 layers of different densities. The neutrino oscillation probabilities are calculated using the OscProb software~\cite{oscprob} where also the non-unitarity neutrino mixing model (see section~\ref{sec:NUNM}) has been implemented. As detailed in~\cite{LoI2016}, secondary particles produced from neutrino interactions with seawater are propagated using GEANT4-based custom KM3NeT software packages; similarly, the light propagation is simulated by taking into account the absorption and scattering lengths as well as the PMT quantum efficiency. Depending on the neutrino energy, the light propagation is either fully simulated or obtained through look-up tables to reduce CPU time consumption.  
The atmospheric muon generation is based on the MUPAGE package~\cite{MUPAGE}, while their propagation in seawater uses a custom software package~\cite{jsirene}. 

%% file: tex/physics.tex
\section{Physics context}
\label{sec:physics}

Due to the very short $\tau$ lepton lifetime, attempting an event-by-event \nutau\xspace identification in KM3NeT/ORCA is currently not possible. Nevertheless, given the atmospheric neutrino flux composition~\cite{flux_2016}, the current active volume, and the GeV-scale sensitivity, the tau neutrino appearance can be quantified on a statistical basis. The $\nu_{\tau}$~normalisation~\cite{SK_tau, opera, IC_tau} is defined as the ratio between the observed and expected number of tau neutrinos. When accounting for the charged-current \nutau~events only, the $\nu_{\tau}$~normalisation can be expressed as a scaling factor of the \nutau~CC cross section as described in section~\ref{sec:cc_ccnc}; therefore, the measurement of the \nutau~normalisation can be interpreted as a direct constraint on the \nutau~CC cross section.

The hypothesis of the unitarity of the neutrino mixing matrix made in atmospheric neutrino experiments constrains the sum of the oscillation probabilities as follows
\begin{equation}
\sum_{\gamma} P_{\beta\gamma}(E,L) = 1
\label{prob_1}
\end{equation}
where $P_{\beta\gamma}$ is the oscillation probability from the neutrino flavour $\beta$ to $\gamma$, $L$ the travelled distance, and $E$ the energy. In this work, the effect of non-unitarity on atmospheric neutrino oscillations, which makes the sum of the measured oscillation probabilities deviate from 1, is studied. The efforts are concentrated on a single parameter that scales the $\tau$-row of the PMNS matrix. This individual parameter affects neutrino oscillation probabilities in the $\nu_{\tau}$ appearance but also in the $\nu_{\mu}$ disappearance channel \cite{Martinez2021, Denton:2021}, as described in section~\ref{sec:NUNM}, enhancing the sensitivity of KM3NeT/ORCA. Section~\ref{sec:NUNM} describes the theoretical framework for this study involving the non-unitarity parametrisation of the neutrino mixing. 

\subsection{From the \nutau~normalisation to the \nutau~CC cross section measurement}
\label{sec:cc_ccnc}

The rate of detected charged-current $\nu_{\tau}/\bar{\nu}_{\tau}$ events can be expressed as
\begin{equation}
\label{PCC0}
N_{\tau} = \int_{E_{\nu}}{\int_{\cos(\theta)}{\sigma^{\text{meas}}_\tau\bigl\{\Phi_e P_{e\tau}+\Phi_\mu P_{\mu\tau}\bigr\}k_N\times\epsilon }}\times dE_{\nu} \times d\cos(\theta),
\end{equation}
where $\sigma^{\text{meas}}_\tau$ is the measured \nutau~cross section, $E_{\nu}$ the neutrino energy, $\theta$ the zenith angle, $k_N$ the number of target nucleons in the detector volume, $\epsilon$ the detector efficiency, and $\Phi_\alpha$ the atmospheric neutrino flux, for any flavour $\alpha$, computed with the Honda model~\cite{honda}. The measured \nutau~cross section $\sigma^{\text{meas}}_\tau$ is related to the theoretical expectation value $\sigma^{\text{th}}_\tau$ used in the Monte Carlo simulation~\cite{genie} through an energy independent scale factor $S_{\tau}$, defined as
\begin{equation}
\label{x-sec}
\sigma_{\tau}^{\text{meas}}(E_{\nu}) = S_{\tau} \times \sigma_{\tau}^{\text{th}}(E_{\nu}).
\end{equation}
$S_{\tau}$ acts as a normalisation, which follows the naming convention used in~\cite{SK_tau}.

\subsection{Non-unitary neutrino mixing}
\label{sec:NUNM}

Beyond Standard Model physics is required to explain the origin of neutrino masses. These may arise from a dimension-5 operator~\cite{CP_2015} that is generated in the seesaw model. It relies on the existence of Heavy Neutral Leptons (HNLs) as seesaw messengers to explain the lightness of neutrino masses compared to other fermions. In the high-scale scenario, where the HNL masses involved are above the electroweak symmetry breaking and beyond experimental reach, the heavy states are kinematically forbidden. In the low-scale scenario, like the inverse or linear seesaw~\cite{InvSS_2009}, the neutrino mixing matrix can be written as a $n\times n$ ($n>3$) unitary matrix. In this scenario, considered here, the new states are kinematically accessible and participate in the neutrino oscillation. The extended mixing matrix remains unitary but $\sum_{\gamma} P_{\beta\gamma}(E,L) \neq 1$ due to the contribution of oscillations involving HNLs.

The non-unitary neutrino mixing (NUNM) can be used to parametrise the effect of $n-3$ HNLs on the neutrino oscillation. This includes the possibility of 3 heavy neutrinos as partners of the standard model (SM) neutrinos, that naturally arise in the type-1 seesaw model. 
The extended $n\times n$ unitary neutrino mixing matrix $U$ can be decomposed as in~\cite{Han2021qum}, in a product of $n(n-1)/2$ rotations:
\begin{equation}
\label{Un}
    U = R_{n-1\,n} \cdot R_{n-2\,n} \cdots R_{3\,n} \cdot R_{2\,n} \cdot R_{1\,n} \cdots R_{2\,3} \cdot R_{1\,3} \cdot R_{1\,2},
\end{equation}
with  $R_{ij}$ representing rotations of angle $\theta_{ij}$ with $i$ and $j$ between $1$ and $n$, and $(R_{2\,3} \cdot R_{1\,3} \cdot R_{1\,2})^{3 \times 3}=U_{\textrm{PMNS}}$ the usual neutrino mixing matrix. When rotations are applied in this specific order, the non-unitary part of $U$ is lower triangular. Based on that, the general structure of the non-unitarity parametrisation is constructed. The lower triangular matrix $\alpha$ is defined as 
\begin{equation}
\label{alpha}
\alpha = \begin{pmatrix}
\alpha_{11} & 0 & 0 \\
\alpha_{21} & \alpha_{22} & 0 \\
\alpha_{31} & \alpha_{32} & \alpha_{33}
\end{pmatrix},
\end{equation}
involving 9 new parameters in addition to the SM mixing angles and phases with real diagonal parameters and complex non-diagonal ones. This lower triangular formalism was first introduced by Okubo in 1962~\cite{Okubo}, in~\cite{Parke_2016, CP_2015, LB_2022}, and tested on data in~\cite{nova_t2k_2019}. The non-unitary matrix $\alpha$ applied to the unitary mixing matrix $U_{\textrm{PMNS}}$ gives the non-unitary matrix $N$ as
\begin{equation}
\label{alpha2}
N = \alpha~ U_{\textrm{PMNS}}.
\end{equation}

This model avoids the computationally intensive parametrisation of these effects with a $n\times n$ unitary mixing matrix in a sterile neutrino model where the number of corresponding angles and phases grows as $n^2$.
The Hamiltonian in matter $H^{n \times n}_{m}$ that describes the evolution of the neutrino state is defined as
\begin{equation}
\label{H0}
H^{n \times n}_{m} = \Delta^{n \times n}  + U^\dagger V^{n \times n}  U
\end{equation}
with $V^{n \times n} =$ diag$\{$V$, 0, ... , 0\}$ and $\Delta^{n \times n} = $ diag\{$\Delta, \Delta m_{41}^2$, ... , $\Delta m_{n1}^2$~ \}. Here, the effective matter potential $V$ that describes the effect of the medium on the neutrino oscillation, and the mass splitting matrix $\Delta$ are defined as 
\begin{equation}
\label{mv}
\Delta = \frac{1}{2E}
\begin{bmatrix}
0 & 0 & 0 \\
0 & \Delta m_{21}^2 & 0 \\
0 & 0 & \Delta m_{31}^2
\end{bmatrix}
\quad
V = 
\begin{bmatrix}
V_{CC} + V_{NC} & 0 & 0 \\
0 & V_{NC} & 0 \\
0 & 0 & V_{NC}
\end{bmatrix},
\end{equation}
with the effective potential due to charged-current and neutral-current interactions $V_{CC} = \sqrt{2}G_Fn_e$ and $ V_{NC} = -\frac{1}{\sqrt{2}}G_Fn_n$, respectively, $G_F$ the Fermi constant, $n_e$ and $n_n$ the electron and neutron number density, respectively.

In this work, the mass of the heavy neutrino $m_{i>3} \gg \sqrt{\Delta m^2_{\text{31}}}$, therefore the model is not sensitive to the HNLs fast oscillations that are averaged out. As discussed in~\cite{nova_t2k_2019}, under this approximation and in the mass basis,
$H^{n \times n}_{m}$ can be truncated into a 3~$\times$~3 matrix, therefore
\begin{equation}
\label{H0bis}
H^{3 \times 3}_{m} = \Delta + N^\dagger V N.
\end{equation}

Only the matter potential proportional to the electron density $V_{CC}$ is relevant in the unitary case; indeed, the matter potential \Vnc,  proportional to the neutron density, affects uniformly all flavours, and is thus equivalent to an absolute phase shift of the neutrino state. Instead, the effect of \Vnc~on active neutrinos only, must be accounted for in the non-unitary case. 

The calculation of the probability after propagation is affected by the non-unitary mixing matrix $N$. $P^{\alpha}_{\beta \gamma}$ describes the oscillation probability to observe flavour $\gamma$ from initial flavour $\beta$ after propagation of a distance $L$ in a fixed density in the presence of unitarity violation (super script $\alpha$) as 
\begin{equation}
\label{P}
P^{\alpha}_{\beta\gamma } = \left| \left( N e^{-i H^{3\times3}_m L} N^\dagger \right)_{\gamma \beta} \right|^2.
\end{equation}
The propagation trough multiple density layers with non-unitary mixing is described in~\cite{fong2023theoretical} and has been implemented in the OscProb software \cite{oscprob} used in this analysis.

The non-observation of very-short-distance oscillations~\cite{nomad, chorus} and searches for light sterile neutrino in atmospheric neutrino experiments, yield the constraints (at 95$\%$ confidence level)  reported in table~\ref{tab:NUparams2} on the parameters of $\alpha$, with $\alpha_{33}$ being the most weakly constrained parameter~\cite{blennow}.

\begin{table}[htbp]
\centering
\begin{tabular}{@{}ccc@{}}
\toprule
\multicolumn{3}{c}{Present bounds } \\ 
\midrule
$\alpha_{33}$ & $>$ & 0.90 \\
$\alpha_{22}$ & $>$ & \xspace0.978 \\ 
$\alpha_{11}$ & $>$ & \xspace0.976 \\
$|\alpha_{32}|$ & $<$ & \xspace0.012 \\
$|\alpha_{31}|$ & $<$ & \xspace0.069 \\
$|\alpha_{21}|$ & $<$ & \xspace0.025 \\
\bottomrule
\end{tabular}
\caption{Current bounds at the 95$\%$ CL derived from~\cite{blennow} for the NUNM parameters.}
\label{tab:NUparams2}
\end{table}

\begin{figure}[htbp]
\centering
\begin{subfigure}{.5\textwidth}
  \centering
  \includegraphics[width=.99\linewidth, trim={0 0 0 0}, clip]{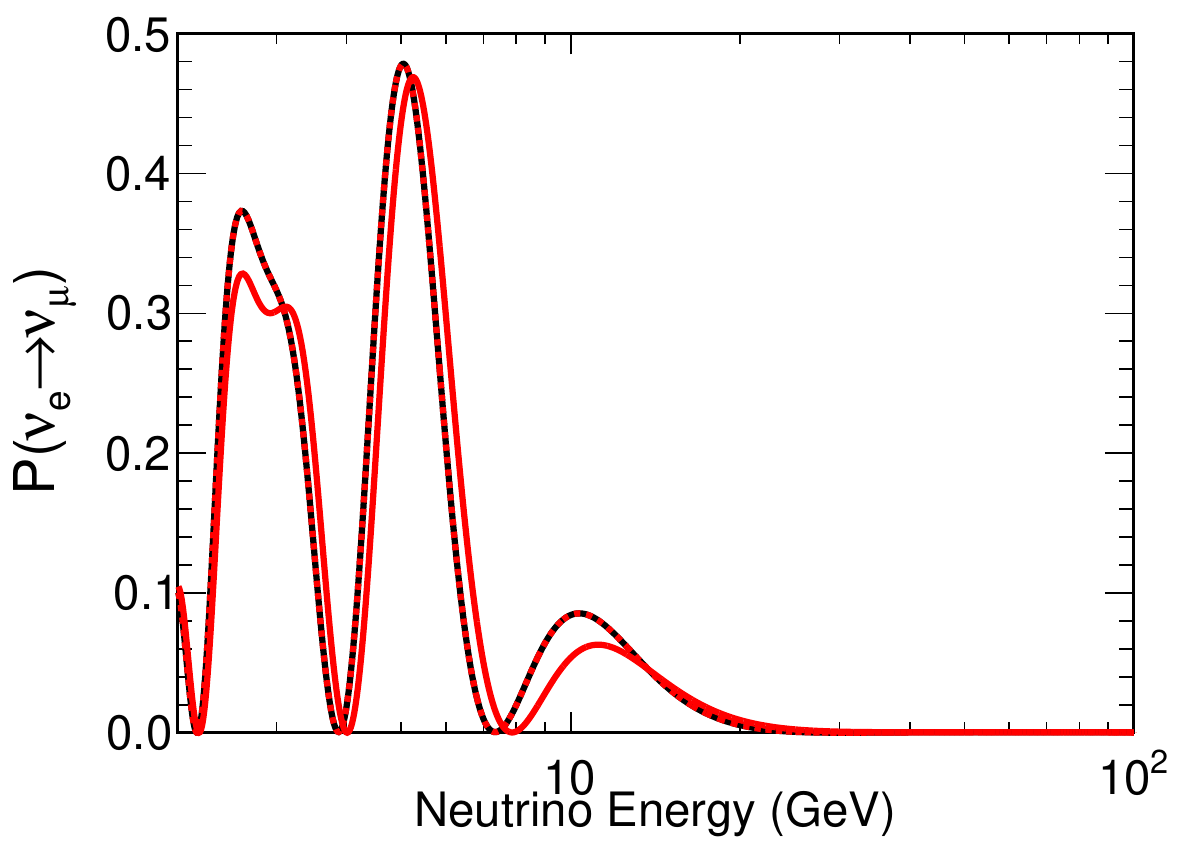}
  \label{fig:sub1}
\end{subfigure}%
\begin{subfigure}{.5\textwidth}
  \centering
  \includegraphics[width=.99\linewidth, trim={0 0 0 0}, clip]{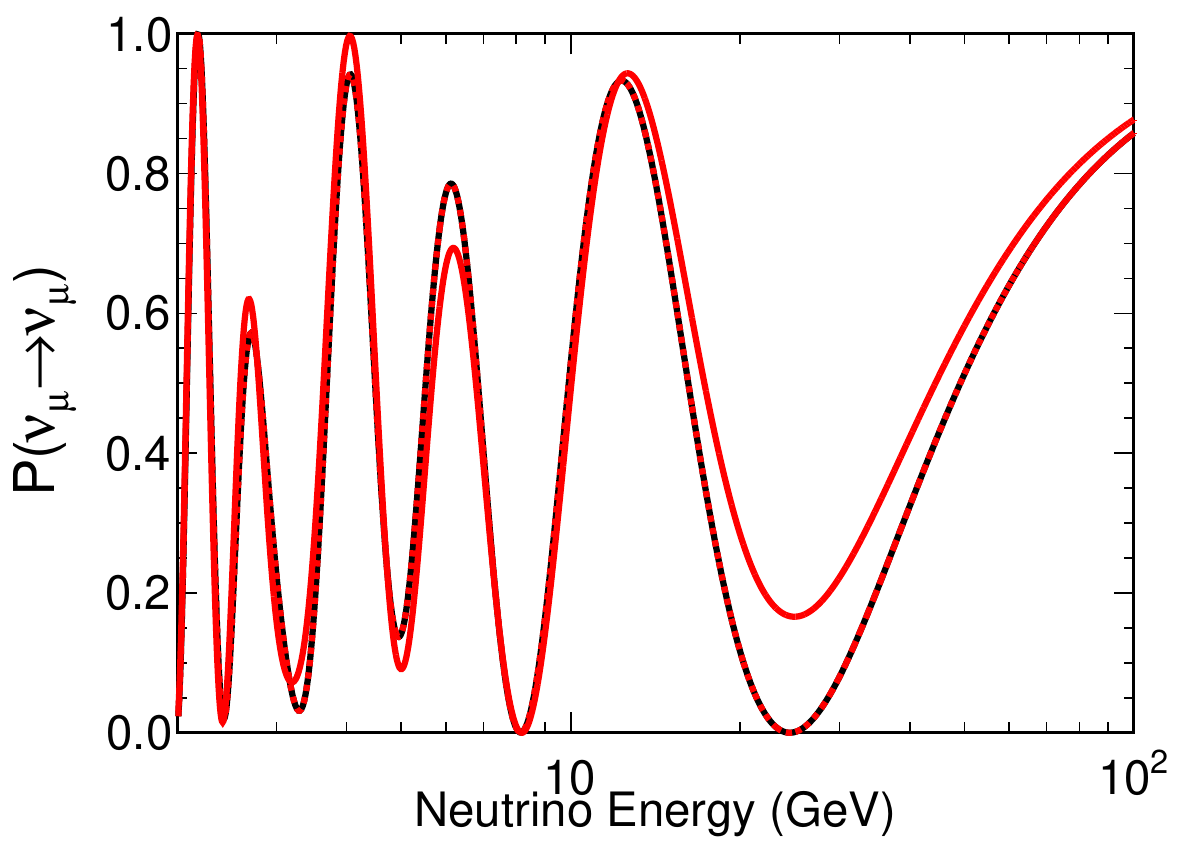} 
  \label{fig:sub2}
\end{subfigure}
\begin{subfigure}{.5\textwidth}
  \centering
  \includegraphics[width=.99\linewidth, trim={0 0 0 0}, clip]{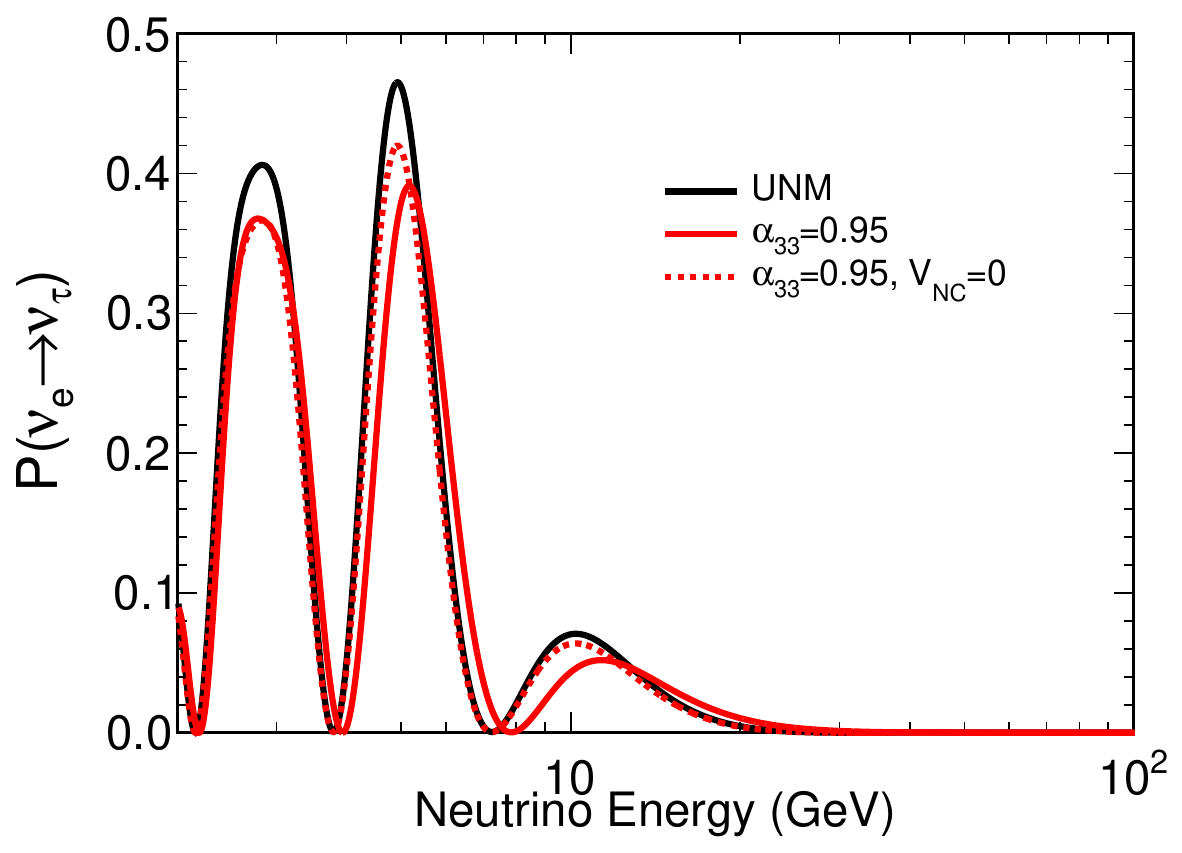} 
  \label{fig:sub3}
\end{subfigure}%
\begin{subfigure}{.5\textwidth}
  \centering
  \includegraphics[width=.99\linewidth, trim={0 0 0 0}, clip]{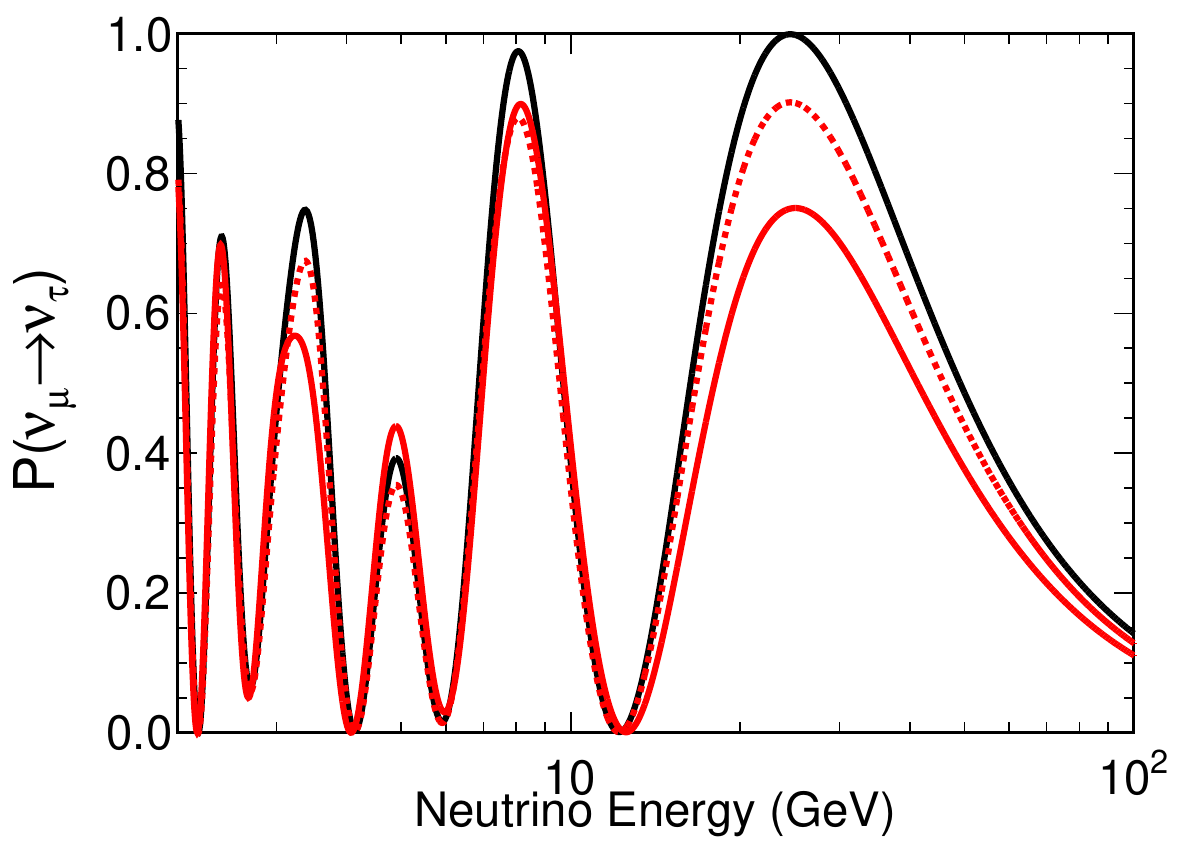}
  \label{fig:sub4}
\end{subfigure}
\caption{The oscillation probabilities are shown for vertically up-going (cos$(\theta)=-1$) neutrinos as a function of neutrino energy for a non-unitary parameter $\alpha_{33}$ being non-zero and for  sin$(2\theta_{23}) = 1$ with and without including the neutral-current potential. The black curve illustrates the Unitary Neutrino Mixing (UNM) case in matter.}
\label{fig:prob_vnc}
\end{figure}

This parameter affects the \nutau~appearance, as it multiplies directly the third row of $U_{\textrm{PMNS}}$ as well as other channels including the $\nu_{\mu}$ disappearance, due to the \Vnc~term. The \Vnc$=0$ case is included in this work, so that the results can be compared to those from~\cite{IC_tau}, which uses the equivalent CC+NC \nutau~normalisation naming convention. In that specific case, the modified probability for a neutrino of initial flavor $\beta$ to remain active is defined as
\begin{equation}
\label{PNC}
P^{\alpha}_{\beta} = P_{\beta e}+P_{\beta \mu}+\alpha_{33}^2P_{\beta \tau}.
\end{equation}
The modified \nutau~CC event rates is given by $N^{\alpha}_{\tau}$ and NC event rates by $N^{\alpha}_{NC}$. They are defined as $N^{\alpha}_{\tau} = \alpha_{33}^2 N_{\tau}$ and by 
\begin{equation}
\label{PNC2}
N^{\alpha}_{NC} = \int_E{\int_{\cos(\theta)}{\sigma_{NC}\Bigl\{\Phi_e[P^{\alpha}_{e}]+\Phi_\mu[P^{\alpha}_{\mu}]\Bigr\} k_N \times \epsilon\times dE \times d\cos\theta}}.
\end{equation}
However, in matter, the equivalence between the non-unitarity framework and the CC+NC approach is broken and the difference in the effect is shown in figure~\ref{fig:prob_vnc}. 
This illustrates that there is a significant contribution of the muon disappearance channel on constraining $\alpha_{33}$, in addition to the tau appearance channel.

By measuring Earth-crossing neutrino oscillations, KM3NeT/ORCA can achieve a competitive sensitivity to $\alpha_{33}$. In addition, when assuming $\cos^2\theta_{34} = \alpha^2_{33}$, this non-unitarity test in the low-scale scenario can be translated into a 1-sterile neutrino case study as described in~\cite{ANTARES_s}, under the hypothesis of $\Delta m_{41}^2 \gg \frac{E}{L}$, $\theta_{14}=\theta_{24}=0$. 

%% file: tex/selection.tex
\section{Event selection}
\label{sec:ev}

The event selection aims at rejecting atmospheric muons and events due to environmental optical background as much as possible and classifying the remaining neutrino candidates according to their topology in a multi-step procedure that was optimised for the measurement of neutrino oscillation parameters. The procedure is summarised hereafter. More details can be found in~\cite{ORCA6} and references therein. 

Events induced by optical background are rejected at first by requiring a minimum number of triggered hits and a good track-reconstruction likelihood. After this first selection, the event sample is mostly composed of atmospheric muons, that can be reduced by selecting only up-going reconstructed tracks, since atmospheric muons from below the horizon are absorbed by the Earth. Finally, a Boosted Decision Tree classifier is used to further discriminate atmospheric muons from atmospheric neutrinos. The remaining events are categorised according to their reconstructed event topology into tracks and showers. The track-like class is split into Low Purity Tracks and High Purity Tracks where purity refers to the contamination of atmospheric muons, that is 4$\%$ and 0.4$\%$, respectively. 

The final dataset contains 5828 events that are distributed across the different classes as shown in table~\ref{tab:event_sel}. In addition, the expected number of events per interaction channel is presented. It is given by the MC event numbers from a fit (see section~\ref{sec:ana}) under the assumption that $S_\tau=1$ and $\alpha_{33}=1$.
According to this model, a total of 185 CC tau neutrino and antineutrino interactions are expected. The majority (77$\%$) are selected in the shower class because $\tau$ leptons produced in \tauCC interactions decay into electrons or hadrons with a total branching ratio of about 83$\%$. The remaining fraction of $\nu_\tau$~CC interactions is evenly distributed between the High Purity Tracks and the Low Purity Tracks.

\begin{table}
	\centering
\begin{tabular}{c c c c c}
	\toprule
	Event Type & Showers & High Purity Tracks & Low Purity Tracks & Total \\ \midrule
	$\nu_e$ CC + $\bar{\nu}_e$ CC & 603 & 51 & 85 & 739 \\
	$\nu_\mu$ CC + $\bar{\nu}_\mu$ CC & 902 & 1777 & 1786 & 4465\\
	$\nu_\tau$ CC + $\bar{\nu}_\tau$ CC & 143 & 22 & 20 & 185 \\
	$\nu$ NC + $\bar{\nu}$ NC & 289 & 13 & 22 & 324\\
	atmospheric $\mu$ + $\bar{\mu}$ & 22 & 7 & 89 & 118\\ \midrule
	Total MC & 1959 & 1870 & 2002 & 5831\\ \midrule
	Total Data & 1958 & 1868 & 2002 & 5828 \\
	\bottomrule
\end{tabular}
\caption{Number of events and composition for the three classes for the MC fitted to the observed data, under the assumption that $\nu_\tau$~normalisation $S_\tau=1$ and non-unitarity parameter $\alpha_{33}=1$ holds. The observed number of events is reported in the last row.}
\label{tab:event_sel} 
\end{table}

%% file: tex/analysis.tex
\section{Analysis method}
\label{sec:ana}
The \nutau~normalisation $S_\tau$ and the non-unitarity parameter $\alpha_{33}$ are measured by fitting a model to the observed event distributions. The event distributions predicted by the model depend on the parameter of interest (PoI), as well as on several nuisance parameters $\epsilon$ that account for systematic uncertainties.

Since neutrino oscillation probabilities depend on the path length $L$ of the neutrino from its production to its detection and on the energy of the neutrino $E$, these quantities are estimated using reconstruction algorithms. The analysis is carried out in the two-dimensional space of reconstructed energy and cosine of the reconstructed zenith angle, which is proportional to the neutrino path length. The two-dimensional distribution is built separately for each of the classes introduced in section~\ref{sec:ev}. Ten equally-spaced bins are used for $\cos\theta$, in the range [$-$1, 0] (where $\cos\theta = -1$ corresponds to vertically up-going neutrinos). For the reconstructed energy, 15 bins between $\SI{2}{GeV}$ and $\SI{1}{TeV}$ are used. The binning is defined in logarithmic scale and optimised such that the expected number of events in each bin is at least 2.

The different models introduced in section~\ref{sec:physics} are fitted to the data through the minimisation of a negative log-likelihood function $-2 \log \mathcal{L}$ defined as:

\begin{equation}
    -2 \log\mathcal{L}  = \sum_{i} 2 \left[  \left( \beta_i N_i^{\text{mod}} - N_i^{\text{dat}} \right) + N_i^{\text{dat}} \ln \left( \frac{ N_i^{\text{dat}} }{\beta_i N_i^{\text{mod}} } \right) \right] + \frac{\left( \beta_i - 1 \right) ^2}{\sigma^2_{\beta_i}} + \sum_k \left( \frac{\epsilon_k - \langle \epsilon_k \rangle}{\sigma_{\epsilon_k}} \right)^2
\label{eq:log-likelihood}
\end{equation}

The sum in the first term of equation~\ref{eq:log-likelihood} runs over each bin $i$ of each of the aforementioned two-dimensional histograms, computing the Poisson likelihood of observing a number of events in data, $N^{\mathrm{dat}}$, given a model with an expectation of $N^{\mathrm{mod}}$. 
The sum in the second term of equation \ref{eq:log-likelihood} runs over the nuisance parameters $\epsilon$ for which constraints have been set by other experiments. This knowledge is taken into account by fitting the systematic uncertainties $\epsilon$ with a Gaussian prior $\sigma_\epsilon$ in order to incorporate it in a consistent manner. The limited Monte Carlo statistics is addressed by introducing normally distributed coefficients $\beta_i$ following the Barlow and Beeston light method~\cite{Barlow:1993dm, Conway:2011in}.
The minimisation is based on the MINUIT library~\cite{James:2296388}.
A detailed description of the nuisance parameters is given in~\cite{ORCA6}, and summarised in table~\ref{tab:parameters}. 
\begin{table}
	\centering		
	\begin{tabular}{c c}
		\toprule
		Parameter & Constraints \\
		\midrule
        $\theta_{23}$ & unconstrained\\
        $\Delta m^2_{31}$ & unconstrained\\
        \midrule
        $f_{\textrm{all}}$ & unconstrained \\
		$f_{\textrm{HPT}}$ & unconstrained \\	
		$f_{\textrm{S}}$ & unconstrained \\
        $f_{\textrm{HE}}$ & 50$\%$ \\		
		$f_{\mu}$ & unconstrained \\
        $f_{\textrm{NC}}$ & 20$\%$ \\
        $s_{\mu\bar{\mu}}$ & 5$\%$ \\
		$s_{e\bar{e}}$ & 7$\%$ \\
		$s_{\mu e}$ & 2$\%$ \\
		$\delta_\gamma$ & 0.3\\
		$\delta_\theta$ & 2$\%$ \\
		$E_s$ & 9$\%$ \\
		$S_\tau$ & depending on analysis \\
		\bottomrule
	\end{tabular}
	\caption{All nuisance parameters and their treatment in the fit. A description of the individual parameters can be found in the text.}
	\label{tab:parameters}
\end{table}
These also include oscillation parameters. The analysis is, however, only sensitive to $\Delta m^2_{31}$ and $\theta_{23}$ which are left unconstrained in the fit, while the other oscillation parameters are fixed to the NuFit v5.0 reference values~\cite{nufit5_0}.
Uncertainties on the shape of the atmospheric neutrino flux are represented by the uncertainty on the spectral index $\delta_\gamma$ and on the slope between horizontal and vertically up-going neutrinos $\delta_\theta$. In addition, the ratios between neutrinos and antineutrinos $s_{\mu\bar{\mu}}$, $s_{e\bar{e}}$ of muon and electron neutrinos respectively, as well as the flavour ratio $s_{\mu e}$ are the uncertainties in the flavour composition of the atmospheric neutrino flux. All systematic uncertainties related to the atmospheric neutrino flux are fitted with constraints considering motivations described in~\cite{Barr_2006}.
A nuisance parameter $f_{\textrm{NC}}$ is introduced to account for uncertainties in the all-flavour neutral-current interaction rate. The PoI for the cross section measurement, $S_\tau$, is an additional nuisance parameter in the non-unitarity fit.
The parameter $f_{\textrm{HE}}$ addresses the differences between the light simulation approaches that have been used in the simulation of low- and high-energy neutrinos, while the energy scale $E_s$ combines several uncertainties associated with the detector response such as PMT efficiencies and water properties extracted from~\cite{LoI2016}. Finally, four normalisation nuisance parameters are considered: overall normalisation $f_{\textrm{all}}$, high-purity track normalisation $f_{\textrm{HPT}}$, shower normalisation $f_{\textrm{S}}$, and atmospheric muon normalisation $f_{\mu}$, allowing for a scaling of the corresponding event numbers.

The sensitivity is estimated by profiling the log-likelihood ratio (LLR)
\begin{equation}
\mathrm{LLR} = -2 \Delta \log\mathcal{L} = \mathrm{min}_{\epsilon} \left(-2 \log\mathcal{L} \right) - \mathrm{min}_{\text{PoI}, \epsilon} \left( -2 \log\mathcal{L} \right)
\end{equation}
where the first term represents the likelihood evaluated using the best-fit values of the nuisance parameters and a fixed value for the PoI, and the second term represents the likelihood evaluated using the best-fit values for the nuisance parameters and for the PoI. The LLR is therefore evaluated for different values of $S_\tau$ or $\alpha_{33}$ to compute the sensitivity and confidence intervals.

The confidence intervals are evaluated with the Feldman-Cousins (FC) method~\cite{Feldman:1997qc} generating 1000 pseudo experiments for several values of the PoI. The nuisance parameters are set to their nominal values assuming normal ordering. Poissonian fluctuations are added to the bin contents in order to account for statistical fluctuations. Finally, the constraints on the nuisance parameters for the fit are randomised by drawing them from a Gaussian distribution according to the mean and uncertainty given in table~\ref{tab:parameters}~\cite{Cranmer:2014lly}. The uncertainty on the limits is obtained by exploiting a bootstrapping technique~\cite{Efron1992}.

%% file: tex/results.tex
\section{Results}
\label{sec:res}

The analysis procedures described in section~\ref{sec:ana} have been applied to fit each of the models described in section~\ref{sec:physics} to the data set described in section~\ref{sec:ev}. The results are summarised in tables~\ref{tab:n_tau} and~\ref{tab:bf_systs_all}, which respectively show the best-fit values for the parameters of interest of each model, and the best-fit values for the nuisance parameters. The PoI is $S_\tau$ when measuring the \tauCC cross section and $\alpha_{33}$ in case of probing non-unitarity.

\begin{table}[hbt]
	\renewcommand*{\arraystretch}{1.2}
	\centering
	\begin{tabular}{c c c c}
		\toprule
		Model & PoI & $N_{\nu_\tau}$ CC & $N_{\nu}$ NC \\
		\midrule
		$S_\tau$ & 0.48$^{+0.5}_{-0.33}$ & 92$^{+90}_{-63}$ & 340$^{+16}_{-16}$\\
		\hline
		$\alpha_{33}$ & 0.993$^{+0.026}_{-0.025}$ & 170$^{+5}_{-9}$ & 325$^{+1}_{-4}$ \\
  	$\alpha_{33}$  ($V_{\textrm{NC}} = 0, S_\tau = 1$)& 0.83$^{+0.20}_{-0.25}$& 132$^{+60}_{-61}$ & 313$^{+11}_{-13}$ \\
		\bottomrule
	\end{tabular}
	\caption{Measured \nutau~normalisation $S_\tau$ and non-unitarity parameter $\alpha_{33}$ and corresponding number $N_{\nu_\tau}$ CC of charged-current tau neutrino and $N_{\nu}$ NC of neutral-current all flavour neutrino interactions for the different considered models. The 1$\sigma$ uncertainties for $S_\tau$ and $\alpha_{33}$ are obtained with the FC approach described in section~\ref{sec:ana}, whereas in case of $\alpha_{33}$ ($V_{\textrm{NC}} = 0, S_\tau = 1$), they are based on Wilks' theorem.}
	\label{tab:n_tau}
\end{table}
\begin{table}
\renewcommand*{\arraystretch}{1.2}
\centering
\begin{tabular}{c c c}
\toprule
& $S_\tau$ & $\alpha_{33}$  \\
Systematic uncertainty& $\epsilon \pm 1 \sigma$ & $\epsilon \pm 1 \sigma$\\
\midrule
$\theta_{23}$ [$^\circ$] & \bfpm{46}{4}{4} & \bfpm{46}{4}{4}\\
$\Delta m^2_{31} [\times 10^{-3}$\,eV$^2$] &  \bfpm{2.15}{0.29}{0.28} & \bfpm{2.18}{0.19}{0.35}\\
\midrule
$f_{\textrm{all}}$ &  \bfpm{1.09}{0.17}{0.11} & \bfpm{1.11}{0.11}{0.12} \\
$f_{\textrm{HPT}}$ & \bfpm{0.91}{0.04}{0.04} &  \bfpm{0.92}{0.04}{0.04} \\	
$f_{\textrm{S}}$ &  \bfpm{0.92}{0.06}{0.06} & \bfpm{0.92}{0.06}{0.06}\\
$f_{\textrm{HE}}$ & \bfpm{1.50}{0.33}{0.30} & \bfpm{1.59}{0.32}{0.29} \\		
$f_{\mu}$ &  \bfpm{0.5}{0.44}{0.4} &  \bfpm{0.51}{0.4}{0.35} \\
$f_{\textrm{NC}}$ & \bfpm{0.89}{0.20}{0.20} & \bfpm{0.86}{0.19}{0.19} \\
$s_{\mu\bar{\mu}}$ &  \bfpm{0.00}{0.05}{0.05} & \bfpm{0.00}{0.05}{0.05} \\
$s_{e\bar{e}}$ & \bfpm{0.01}{0.07}{0.07} & \bfpm{0.01}{0.07}{0.07} \\
$s_{\mu e}$ & \bfpm{-0.004}{0.020}{0.020} & \bfpm{-0.004}{0.020}{0.019}  \\
$\delta_\gamma$ & \bfpm{-0.019}{0.027}{0.026} & \bfpm{-0.019}{0.025}{0.025}\\
$\delta_\theta$ & \bfpm{-0.002}{0.019}{0.019} & \bfpm{-0.005}{0.019}{0.019} \\ 
$E_s$ & \bfpm{0.98}{0.11}{0.08} & \bfpm{1.03}{0.08}{0.11} \\
$S_\tau$ & PoI & \bfpm{0.92}{0.19}{0.18}\\
\bottomrule
\end{tabular}
\caption{All systematic uncertainties and their best-fit values along with their 1$\sigma$ post-fit uncertainties. The description of the parameters is given in the text.}
\label{tab:bf_systs_all}
\end{table}

\subsection{$\nu_{\tau}$ CC cross section measurement}
\label{sec:cc}

\begin{figure}[h!tpb]
\centering
\includegraphics[width=0.7\columnwidth]{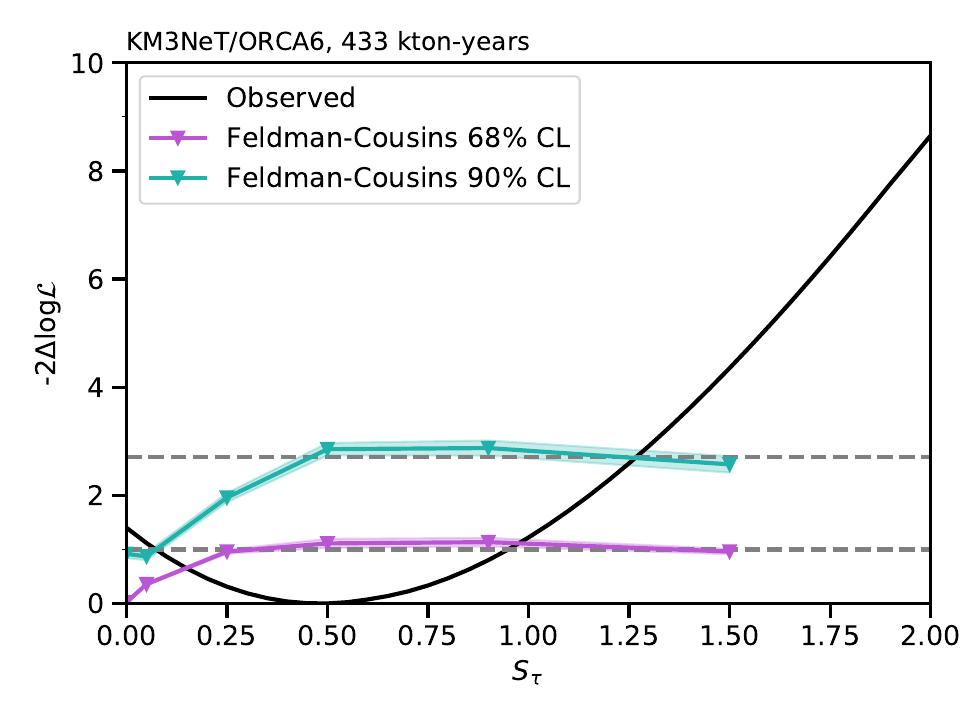}
 \caption{Measured log-likelihood profile with Feldman-Cousins correction for the model probing the \tauCC cross section (black line). The horizontal dashed lines represent the 68\% and 90\% CL thresholds according to Wilks' theorem, whereas the coloured lines indicate the Feldman-Cousins correction with the corresponding 1$\sigma$ uncertainty.}
 \label{fig:profile_cc}
\end{figure}

The parameter $S_\tau$ and the corresponding 1$\sigma$ uncertainty were measured to be 0.48$^{+0.5}_{-0.33}$. The best fit corresponds to a total number of 92$^{+90}_{-63}$ observed \tauCC events and a total of 340$^{+16}_{-16}$ NC interactions. The profiled log-likelihood for $S_\tau$ reported in figure~\ref{fig:profile_cc} allows for the visualisation of the best-fit value as well as the FC correction of the \SI{68}{\%} and \SI{90}{\%} confidence intervals.
The dashed lines indicate the 68\% and 90\% CLs according to Wilks' theorem~\cite{Wilks:1938dza}, whereas the coloured lines represent the CLs from the FC correction with their corresponding 1$\sigma$ uncertainty derived from bootstrapping. As expected, the CLs from Wilks' theorem and FC mostly deviate from each other at low values of $S_\tau$, i.e.\ close to the physical boundary. Above $S_\tau=0.25$ (0.50) the difference for the 68\% (90\%) CL is negligible.
The p-value to exclude the hypothesis of $S_\tau = 0$ is $(5.9\pm 0.8)\%$, calculated as the fraction of pseudo datasets for which $-2 \Delta \log\mathcal{L}$ is larger than the observed one.

\begin{figure}[h!tpb]
\centering
\includegraphics[width=0.7\columnwidth]{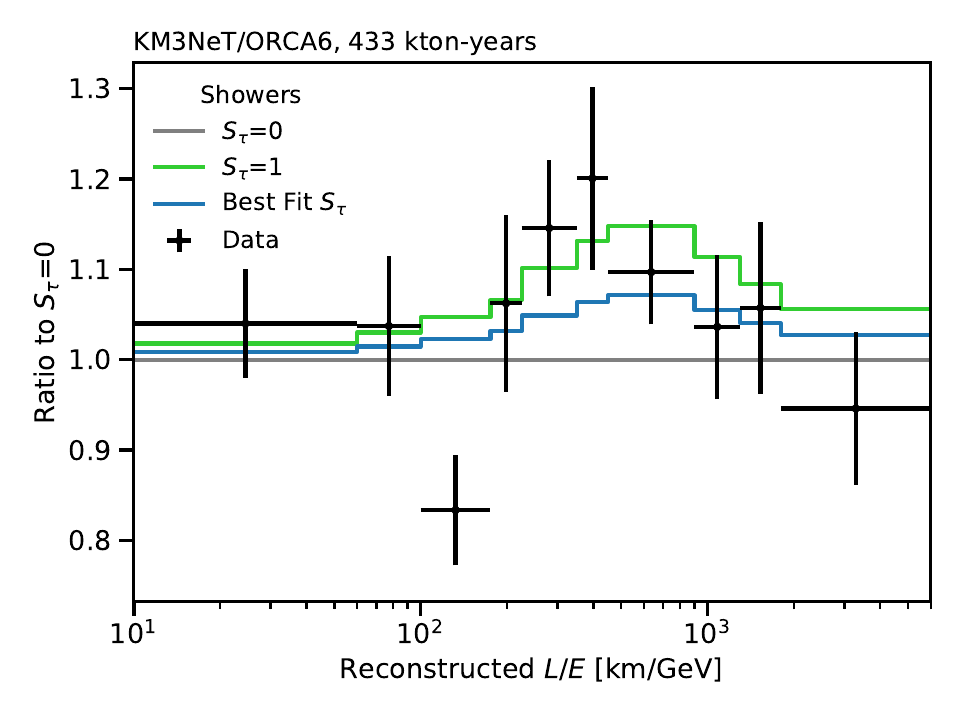}
 \caption{Measured $L/E$ distribution for the shower class (black crosses) with best-fit model (blue line) and a model with nominal \nutau~normalisation $S_\tau$ (green line) with respect to no-\nutau~appearance. }
 \label{fig:loe_cc}
\end{figure}

The $L/E$ distribution is shown in figure \ref{fig:loe_cc}. Even though it is not used in the analysis, this distribution helps to visualise the results since the oscillation probability is a function of $L/E$.
The ratio for the observed data compared to the best-fit model and a model with $S_\tau = 1$ to a model with the extreme case $S_\tau = 0$ is presented. The distribution is shown for the shower class, that is the most sensitive class to the observation of \nutau~appearance.

\begin{figure}[h!tbp]
	\centering
	\centering
	\includegraphics[width=.9\linewidth, trim={0 0 0 0}, clip]{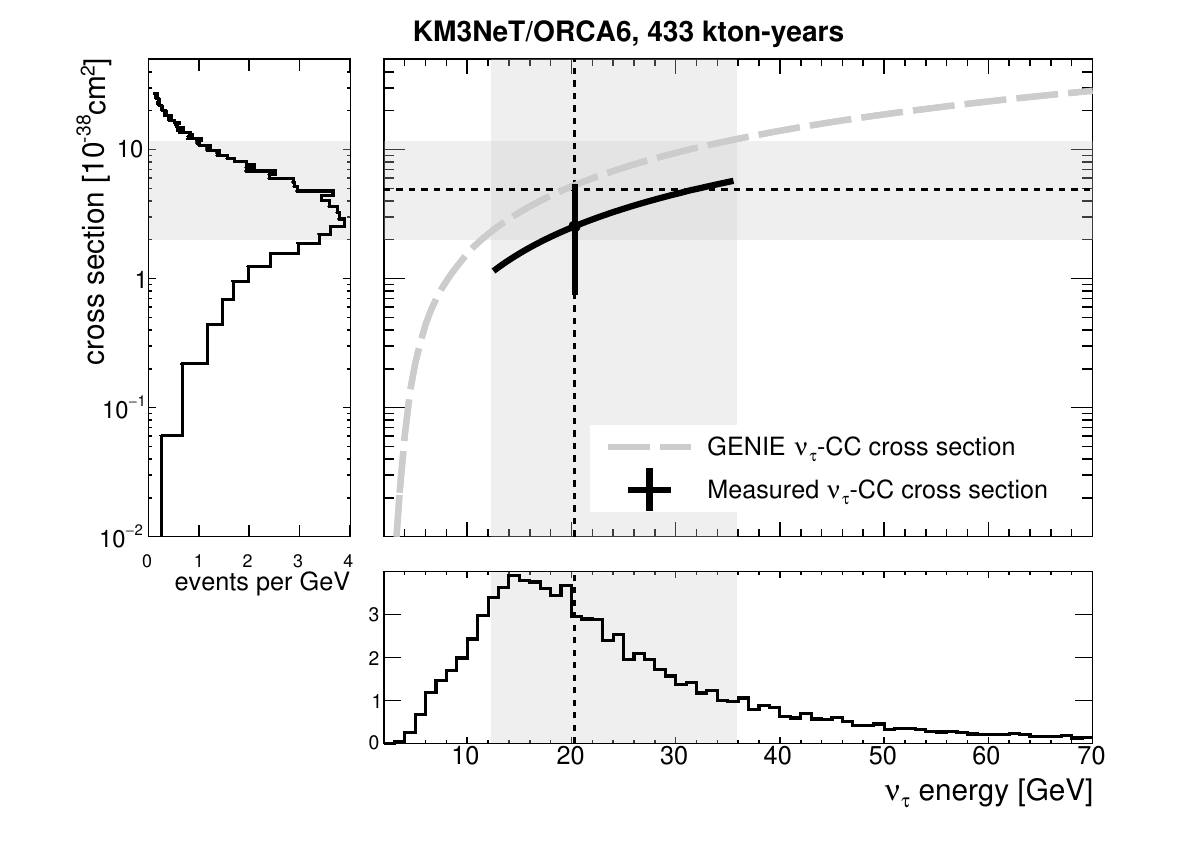} 
	\caption{Measurement of the charged-current tau neutrino cross section in black, compared to the $\nu_{\tau}/\bar{\nu}_{\tau}$ averaged theoretical expectation in grey as a function of the energy. The light grey bands represent the 68\% range of the distributions.}
	\label{fig:x-sec}
\end{figure}

The measured $\nu_\tau$~CC cross section is related to the theoretical expectation following equation~\ref{x-sec}. In figure~\ref{fig:x-sec}, the energy-dependent theoretical expectation from~\cite{genie} is scaled using the best-fit value $S_{\tau} = (0.48 ^{+0.5}_{-0.33})$, 
while the median of the $\nu_\tau$~CC true energy distribution $E_{\nu_{\tau}}$ is at 20.3 GeV between 12.3 and 35.9 GeV at 68\% CL. The corresponding theoretical expectation is $\sigma_{\tau}^{\text{th}} = 5.29 \times 10^{-38} $ cm$^{2}$, calculated as the average of the neutrino and antineutrino cross section according to equation~\ref{eq:x-sec_average}
\begin{equation}
	\begin{split}
		\sigma_{\tau}^{\text{th}} &= r_\nu \sigma_{\tau} + (1-r_\nu)\sigma_{\bar{\tau}} \\
		r_\nu &= X_\nu / (X_\nu + \lambda X_{\bar{\nu}}) \\
		X_\nu &= \int_{E_\nu} \int_{\cos(\theta)} \sigma_{\tau} \bigl\{ \Phi_e P_{e\tau} + \Phi_\mu P_{\mu\tau}\bigr\} \times dE_{\nu} \times d\cos(\theta)\\
		X_{\bar{\nu}} &= \int_{E_{\bar{\nu}}} \int_{\cos(\theta)} \sigma_{\bar{\tau}} \bigl\{ \Phi_{\bar{e}} P_{\bar{e}\bar{\tau}} + \Phi_{\bar{\mu}} P_{\bar{\mu}\bar{\tau}}\bigr\} \times dE_{\bar{\nu}} \times d\cos(\theta).
	\end{split}
	\label{eq:x-sec_average}
\end{equation}
The neutrino fraction is found as $r_\nu = 0.72$. $X_\nu$ and $X_{\bar{\nu}}$ represent --- up to some constants which cancel in the ratio --- the neutrino and antineutrino interaction rate while $\lambda$ accounts for potential differences in the reconstruction efficiencies between neutrino and antineutrino events due to their non-identical interaction inelasticities. However, the invisible neutrino from the tau decay fully compensates for the presumed inelasticity effect and $\lambda = 1.0$ is derived.
Hence, the measured cross section is $\sigma_{\tau}^{\text{meas}} = (2.5 ^{+2.6}_{-1.8}) \times 10^{-38} $ cm$^{2}$.

\begin{figure}[h!tpb]
    \centering
	\includegraphics[width=0.7\columnwidth]{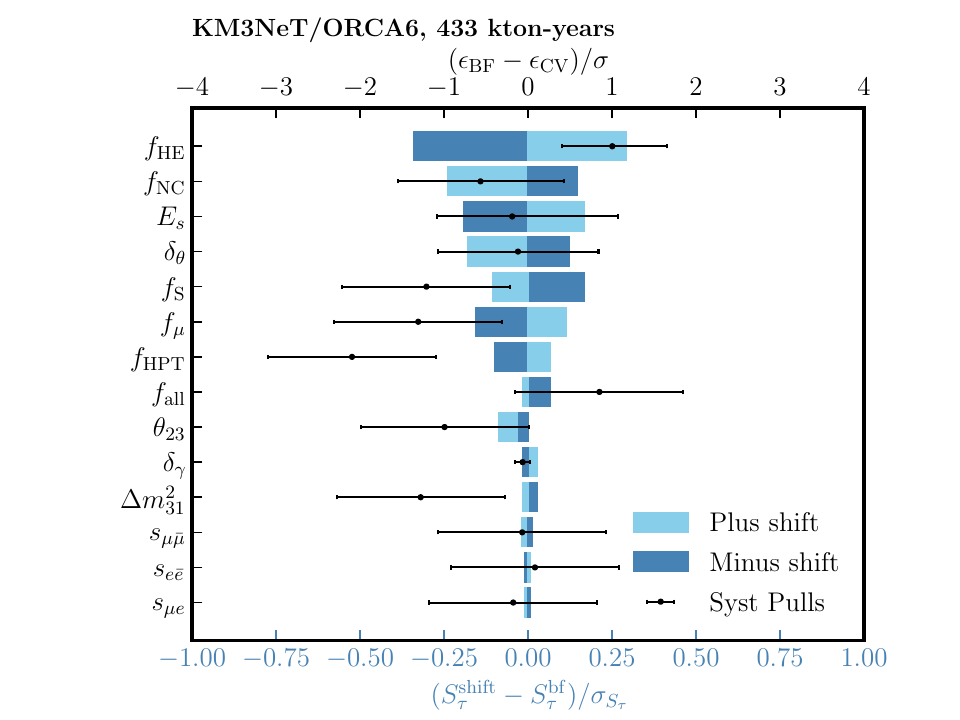}
    \caption{Impact on the best-fit value of $S_\tau$ when shifting a systematic uncertainty by plus (light blue) or minus (dark blue) the 1$\sigma$ post-fit uncertainty from its best-fit value (lower $x$-axis). The pulls (black markers, upper $x$-axis) represent the deviation of the best-fit value from the central value of the parameter with respect to its uncertainty.}
	\label{fig:pulls_cc}
\end{figure}

The effect of the nuisance parameters on the fit result of $S_\tau$ is presented in figure~\ref{fig:pulls_cc}, where blue bars represent the impact of a nuisance parameter on $S_\tau$. Each of them is calculated by performing two fits: one in which the nuisance parameter is fixed to the best-fit value from table~\ref{tab:bf_systs_all} plus its 1$\sigma$ post-fit uncertainty and one in which it is fixed to the best-fit value minus its uncertainty. Each fit yields a new fitted value for $S_\tau$ denoted as $S_\tau^{\textrm{shift}}$, which allows for the calculation of the impact as $\left(S_\tau^{\mathrm{shift}} - S_\tau^{\mathrm{bf}}\right) / \sigma_{S_\tau}$, where $S_\tau^{\textrm{bf}}$ is the best-fit value of $S_\tau$, and $\sigma_{S_\tau}$ its 1$\sigma$ error. This plot also shows the pulls for each nuisance parameter (black markers), which are defined as the difference between the best-fit value $\epsilon_{\textrm{BF}}$ and the expected value $\epsilon_{\textrm{CV}}$ of the corresponding parameter, divided by its uncertainty $\left(\epsilon_{\textrm{BF}} - \epsilon_{\textrm{CV}} \right) / \sigma$. If available, $\sigma$ is given by the pre-fit uncertainty (see table~\ref{tab:parameters}), and the error bars represent the ratio of the post-fit and pre-fit uncertainties $\sigma_{\epsilon}^{\mathrm{post-fit}} / \sigma_{\epsilon}^{\mathrm{pre-fit}}$. For the parameters that are fitted without constraints, $\sigma$ is given by the post-fit uncertainty, and the error bars are set to 1.

The high-energy light simulation $f_{\textrm{HE}}$ and the $f_{\textrm{NC}}$ normalisation are the nuisance parameters with the largest impact on $S_\tau$.
The energy scale $E_s$, i.e.\ the detector-related uncertainties, has a non-negligible impact. Regarding the atmospheric neutrino flux parameters, only the slope between up- and horizontally-going neutrinos $\delta_\theta$ has a sizeable influence, whereas the three composition ratios do not affect the $S_\tau$ measurement significantly. Systematic uncertainties that are fitted without constraints have larger pulls, as expected. Among the systematic uncertainties that are fitted with a constraint, only $f_{\textrm{HE}}$ and $f_{\textrm{NC}}$ show large pulls. Except for $\delta_\gamma$, for which the post-fit uncertainty is significantly reduced compared to the pre-fit one, all other nuisance parameters have a negligible improvement in their post-fit uncertainty.

\subsection{Probing unitarity}
\label{sec:nc}

\begin{figure}[h!tpb]
\centering
\includegraphics[width=0.49\textwidth]{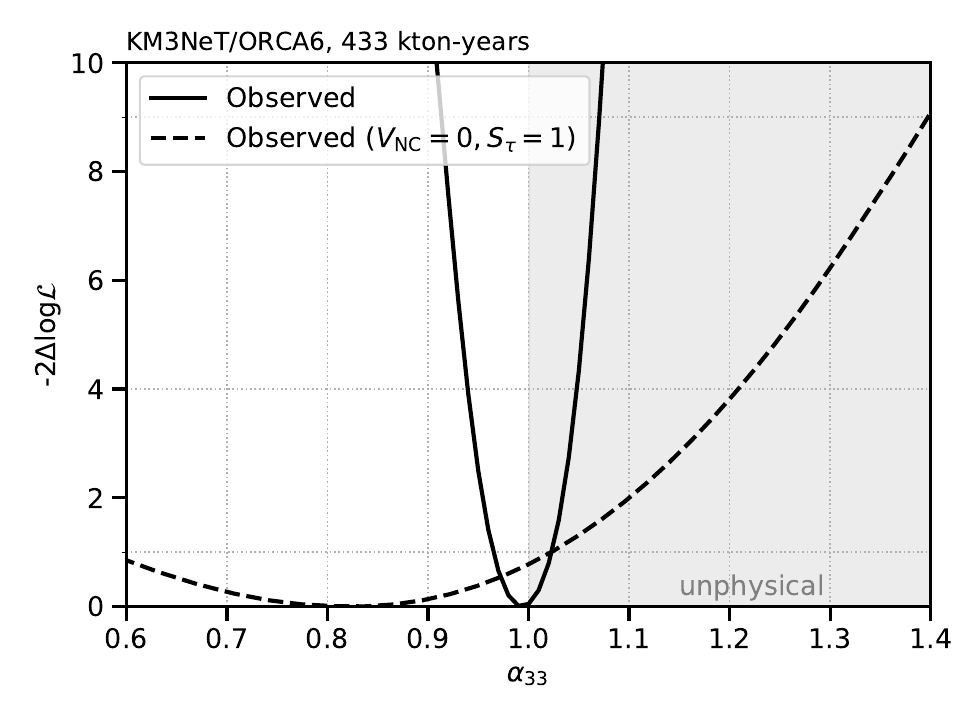}
\includegraphics[width=0.49\textwidth]{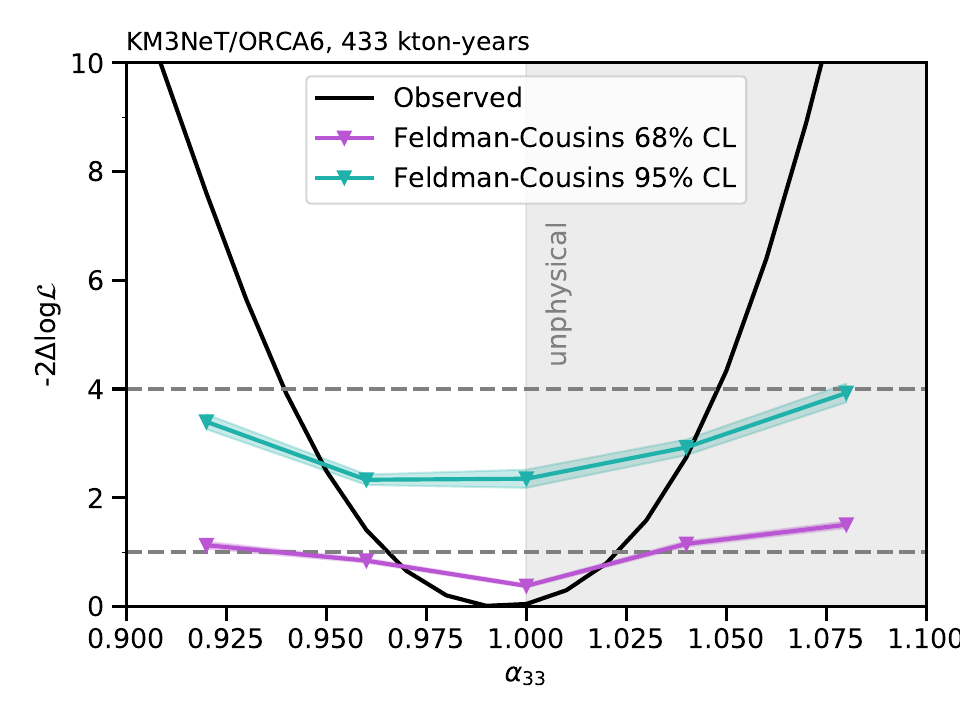}
\vspace{-6.5pt}\small{\caption{\label{fig:profile_nunm} Measured log-likelihood profile for the model probing non-unitarity (black solid lines). \textbf{Left:} Comparison with a model with $V_{\textrm{NC}}=0$ and $S_\tau = 1$ (black dashed line). \textbf{Right:} Feldman-Cousins correction. The horizontal dashed lines represent the 68\% and 95\% CL thresholds according to Wilks' theorem, whereas the coloured lines indicate the Feldman-Cousins correction with the corresponding 1$\sigma$ uncertainty.}}
\end{figure}

The scale factor for the $\nu_\tau$~CC cross section $S_\tau$ is included as a nuisance parameter for the measurement of $\alpha_{33}$, and assigned a \SI{20}{\%} uncertainty. The best-fit value is ${\alpha_{33}=0.993^{+0.026}_{-0.025}}$ corresponding to 170$^{+5}_{-9}$ \tauCC and a total of 325$^{+1}_{-4}$ NC events. The profiled log-likelihood ratio is presented in figure~\ref{fig:profile_nunm}. On the left hand side, it is compared to a model without neutral-current potential, i.e.\ $V_{\textrm{NC}}=0$, and with $S_\tau = 1$ for which the best-fit value is $\alpha_{33}=0.83^{+0.20}_{-0.25}$, corresponding to 132$^{+60}_{-61}$ \tauCC and a total of 313$^{+11}_{-13}$ NC events. The latter results can be compared to the results published by the IceCube collaboration~\cite{ IC_tau} following the physics interpretation from section~\ref{sec:physics}, where the CC+NC $\nu_\tau$~normalisation corresponds to $\alpha_{33}^2$ with $V_{\textrm{NC}} = 0$. The difference in the sensitivities for both models originates from the fact that when $V_{\textrm{NC}}=0$, $\alpha_{33}$ affects only the \nutau~appearance probability, as shown in figure~\ref{fig:prob_vnc}. If on the other hand $V_{\textrm{NC}} \neq 0$, $\alpha_{33}$ additionally affects other oscillation channels including $\nu_\mu$~disappearance leading to better constraints (as discussed in section~\ref{sec:conc}).

The FC correction is shown on the right side of figure~\ref{fig:profile_nunm}. Due to the high computational cost, it is only applied for the case where $S_\tau$ is fitted and $V_\textrm{NC}$ is included. The results are similar to Wilks’ theorem except for values of $\alpha_{33}$ that are close to 1, which suggests that there is a physical boundary. In fact, $\alpha_{33} > 1$ is an unphysical region for which the sum of oscillation probabilities would exceed 1. Overall, the 68$\%$ confidence interval is almost unaffected, whereas the 95$\%$ confidence interval is slightly reduced with $\alpha_{33} >$ 0.95 at 95$\%$ CL.

The best fit is consistent with unitarity, i.e.\ $\alpha_{33} = 1$ (p-value of $68\% \pm 2\%$), also when neglecting the $V_\textrm{NC}$ potential and fixing the \tauCC cross section (0.6$\sigma$ according to Wilks’ theorem).

\begin{figure}[h!tpb]
\centering
\includegraphics[width=0.7\columnwidth]{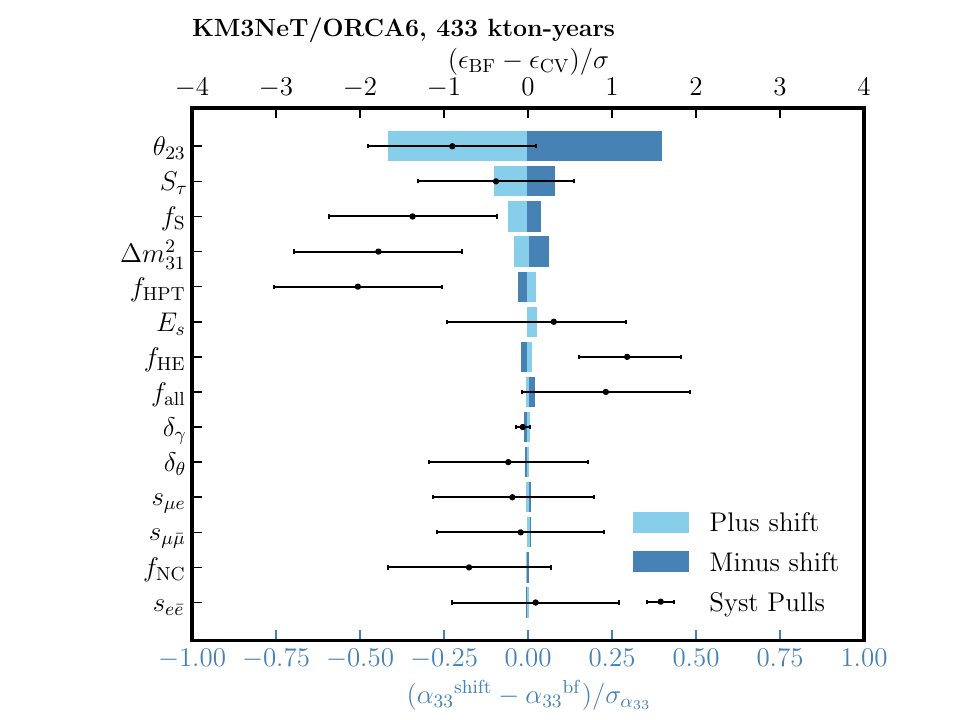}
\caption{Impact on the best-fit value of $\alpha_{33}$ when shifting a systematic uncertainty by plus (light blue) or minus (dark blue) the 1$\sigma$ post-fit uncertainty from its best-fit value (lower $x$-axis). The pulls (black markers, upper $x$-axis) represent the deviation of the best-fit value from the central value of the parameter with respect to its uncertainty.}
\label{fig:pulls_nc}
\end{figure}

Impacts and pulls for the measurement of $\alpha_{33}$ are presented in figure~\ref{fig:pulls_nc}. Overall, the pulls of the nuisance parameters are comparable with what is obtained from the measurement of $S_\tau$, except for $E_s$ which has the opposite sign and a larger pull for $\Delta m_{31}^2$. The nuisance parameter with the largest impact on $\alpha_{33}$ is clearly $\theta_{23}$. 
This can be explained by the fact that both parameters affect the $\nu_\mu$ disappearance. Additionally, $\alpha_{33}$ has an influence on the \nutau~appearance channel which can partly be compensated by $S_\tau$ leading to a non-negligible impact of this parameter. Apart from $\Delta m_{31}^2$ and $f_\textrm{S}$, the remaining nuisance parameters do not have a major impact on the result.

%% file: tex/conclusions.tex
\section{Comparison with other experiments and conclusions}
\label{sec:conc}

Exploring ORCA6 data, two different analyses for the \nutau~normalisation and a test of the unitarity of the neutrino mixing matrix have been presented.
A data sample of 433~kton-years exposure collected with 5$\%$ of the nominal instrumented volume (corresponding to 6 operational detection units) has been used for these analyses. 

\begin{figure}[htpb]
\centering
\includegraphics[width=7.4cm, height=4.4cm]{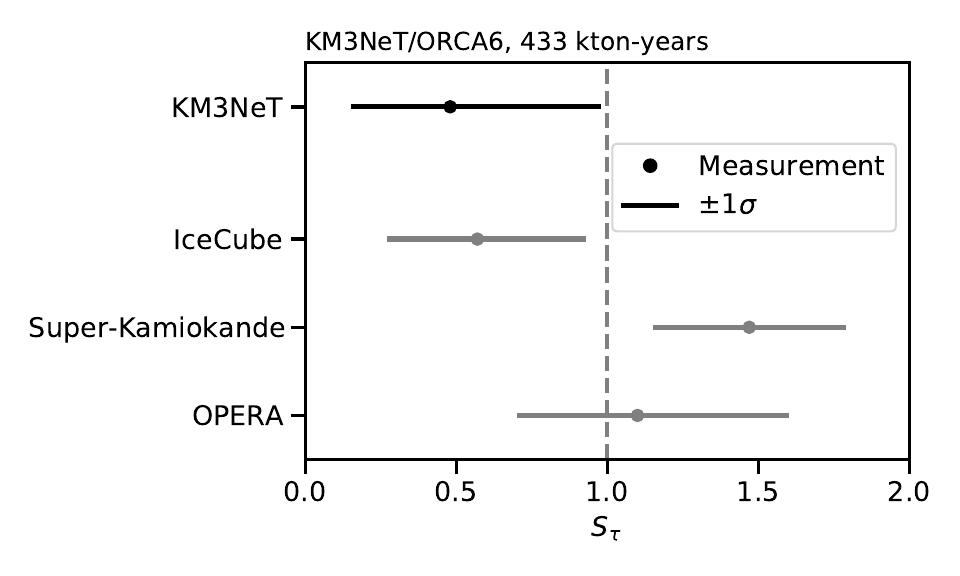}
\includegraphics[width=7.4cm, height=4.4cm]{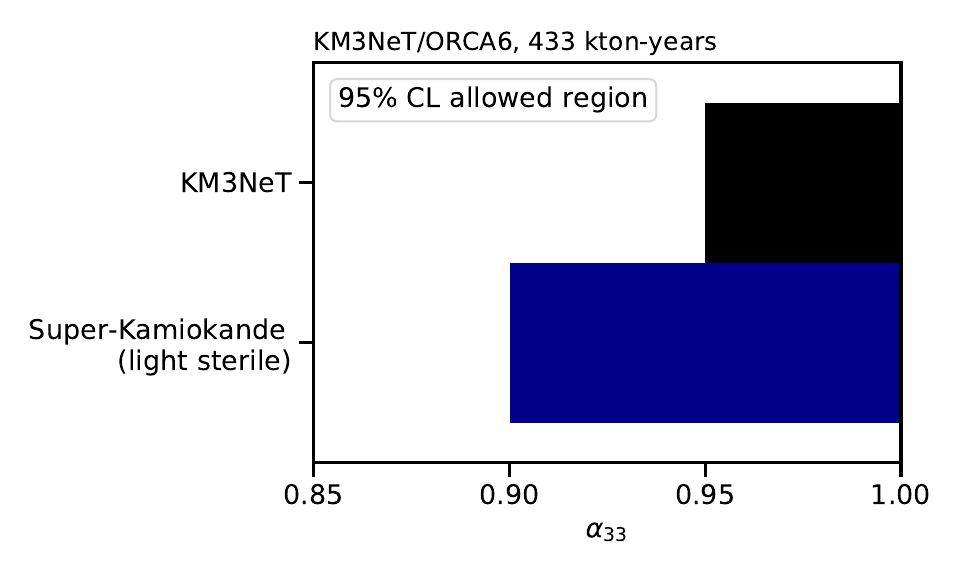}
\vspace{-6.5pt}\small{\caption{\label{fig:comparison_cc_nc} \textbf{Left:} comparison of \Stau\xspace measurement of \osix\xspace with previous results on \nutau~CC cross section normalisation reported by OPERA~\cite{opera2018}, Super-Kamiokande~\cite{SK_tau}, and IceCube~\cite{IC_tau}. \textbf{Right:} comparison of the $\alpha_{33}$ constraints of \osix\xspace with the measurement from Super-Kamiokande derived in~\cite{blennow}. 
}} 
\end{figure}

The first analysis explores the \nutau~appearance channel for a measurement of the \nutau~normalisation in the oscillation channel and explicitly assumes that the \threeflav\xspace is valid. Consequently, possible deviations of the \nutau~normalisation from 1 correspond to a scaling of the \nutau~CC event rate appearing from the oscillated atmospheric neutrino flux. This method is the same as the one used by other experiments exploring the \nutau~appearance oscillation either in the $\nu_{\mu}$ CNGS beam (as in OPERA~\cite{opera}) or in the atmospheric flux (as in Super-Kamiokande~\cite{SK_tau} and IceCube~\cite{IC_tau}). Nonetheless, the comparison among previously reported measurements is not straightforward because it is based on the detection of neutrinos oscillating at various neutrino energies and interacting in different media. Moreover, they have complementary sensitivity to the oscillation parameters; consequently, their treatment of the \nutau~normalisation fit is different. 
As in the atmospheric neutrino experiments Super-Kamiokande and IceCube, the ORCA6 analysis is performed on a statistical basis at the oscillation maximum around 25\,GeV, due to the impossibility of directly tagging \nutau\xspace and $\bar{\nu}_{\tau}$ events. In the ORCA6 event sample, 92$^{+90}_{-63}$ \nutau~CC are observed. The \nutau~normalisation is fitted while keeping the $\theta_{23}$ angle and the $\Delta m^2_{31}$ mass splitting parameters free. Both normal and inverted ordering hypotheses as well as both $\theta_{23}$ octants are tested. In this way, the \nutau~normalisation is measured at 0.48$^{+0.5}_{-0.33}$. The result is reported in terms of the significance of excluding the hypothesis $S_{\tau}$~=~0, corresponding to a p-value of (5.9 $\pm$ 0.8)~$\%$. 
The up to date overview of the \nutau~CC normalisation measurements at 68$\%$ CL, including the \osix\xspace result, is shown in the left plot of figure~\ref{fig:comparison_cc_nc}. 
None of the reported results shows statistical evidence of rejecting the standard \threeflav\xspace description. The achieved ORCA6 precision is already in line with other reported measurements, despite the reduced exposure of this analysis.  

Under the hypothesis that the PMNS matrix is unitary, the \nutau~normalisation \Stau can be interpreted as a scale factor of the \nutau~CC cross section, allowing to constrain its measurement. Up to now, the uncertainty on the cross section is still one of the main sources of systematics in neutrino oscillation experiments, and measurements of the \nutau~CC cross section have been reported only by a few experiments. Hence, the diverse energy range and interaction media in the various experiments as well as the different impacts of systematic effects and uncertainties is an additional motivation for pursuing such measurements. In order to measure \nutau~CC cross section in \osix\xspace data, the theoretical \nutau\xspace cross section at the median of the \nutau\xspace energy distribution, 20.3\,GeV, within [12.3, 35.9]\,GeV at 68\% CL, is considered. 
In the case of OPERA~\cite{opera2018}, the accessible neutrino energy range is restricted below 20\,GeV and \nutau~CC interactions take place in a lead-based target. In experiments exploiting the \nutau~appearance channel from the atmospheric neutrino oscillation, the separation between $\nu_{\tau}$ and $\bar{\nu}_{\tau}$ cannot be distinguished.
Similarly to Super-Kamiokande, water is the interaction medium also for the ORCA6 measurement. Here, the $\nu_\tau$ CC cross section is found to be $\sigma_{\text{meas}} = (2.5 ^{+2.6}_{-1.8}) \times 10^{-38} $ cm$^{2}$ and, as in the case of Super-Kamiokande, is still dominated by statistical uncertainties.
This result is in good agreement with what was reported by OPERA. For a better comparison highlighting the difference among the experiments mentioned above, a summary of the \nutau~CC cross section measurements and the corresponding statistics is reported in table~\ref{tab:nutauccxsec_comparison}.

\begin{table}[htpb]
    \centering
    \begin{tabular}{p{3.9cm}|p{1.8cm}|p{1.8cm}|p{2.2cm}|p{3.2cm}}
        \hline
        Experiment & Interaction  & Energy & N. of  & $\nu_{\tau}$ CC cross\\
                   &  medium      & [Gev]  & observed \nutau\xspace & section [nucleon$^{-1}$\\
                   &              &        &                        & 10$^{-38}$\,cm$^2$] \\
        \hline
        \hline
        \vspace{-12pt} & & & \\
        OPERA~\cite{opera2018} & lead & $\leq$20 & 10 & 2.46$^{+1.15}_{-0.98}$ \\
        Super-Kamiokande~\cite{SK_tau} & water & $\sim$25 & 338.1$\pm$72.7 & 0.94 $\pm$ 0.20 \\
        ORCA6 (this work) & water & 20.3$^{-8.0}_{+15.6}$ & 92$^{+90}_{-63}$ & 2.5$^{+2.6}_{-1.8 }$ \vspace{3pt}\\
        \vspace{-21pt} & & & \\
        \hline
    \end{tabular}
    \caption{Summary of the reported measurements of the \nutau~CC cross section and the corresponding statistics; the measurement from OPERA~\cite{opera2018} has been converted to the same units as the results by Super-Kamiokande and ORCA6, dividing by the number of nucleons, depending on the target composition.}
    \label{tab:nutauccxsec_comparison}
\end{table}

The second major result obtained from exploring this data sample is the explicit non-unitarity test of the neutrino mixing matrix, with the purpose of reaching a higher precision in validating the \threeflav. In the analysis described in this paper, the extension to a general $n \times n$ neutrino mixing matrix has been considered, by using the formalism described in section~\ref{sec:NUNM}. Following this approach, the small non-zero neutrino masses are further suppressed by the existence of Heavy Neutral Leptons participating in the neutrino oscillations. Hence, the non-unitarity of the PMNS matrix is tested by extending it with with additional parameters $\alpha_{ij}$. A precise measurement of these extra parameters has been performed only partially on data in the accessible energy range of long-baseline neutrino experiments at accelerators~\cite{nova_t2k_2019}. The ORCA6 result is the first one obtained with Earth-crossing atmospheric neutrinos. By exploring this neutrino sample, the best sensitivity on the most weakly constrained $\alpha_{33}$ parameter is accessible due to matter effects and the non-negligible contribution of the \Vnc~potential to oscillation probability. 
The fit of the $\alpha_{33}$ parameter is performed under two different assumptions: the most general one where \Stau\xspace is considered as a
nuisance parameter with 20~$\%$ assigned uncertainty, and the specific case with \Stau=1 and $V_{\textrm{NC}}$=0. The latter corresponds to a description where optimal knowledge on the \nutau~CC cross section is assumed and the neutron density does not contribute to the matter potential. This restricted description of the NUNM formalism is considered in the ORCA6 analysis for a direct comparison with the CC+NC normalisation reported by the IceCube Collaboration~\cite{IC_tau}, under the PMNS unitarity hypothesis, where a variation in both CC and NC rates would affect the \nutau~normalisation measurement. In ORCA6, $\alpha_{33}$ is found at 0.83$^{+0.20}_{-0.25}$; both the \nutau~CC and NC rates vary so that 132$^{+60}_{-61}$ \nutau~CC and 313$^{+11}_{-13}$ NC events are counted, respectively. This result is consistent with the IceCube CC+NC \nutau~normalisation measurement, and comparable precision is achieved in both analyses.

However, as described in  section~\ref{sec:physics}, the \Vnc~potential cannot be neglected in matter, given that it affects all the oscillation channels; hence, the $\alpha_{33}$ parameter is fitted including the \Vnc~potential and allowing the \Stau~parameter to vary. Under these hypotheses, $\alpha_{33}$ is 0.993$^{+0.026}_{-0.025}$. The reported ORCA6 limit on the $\alpha_{33}$ parameter, reduced at [0.95, 1.04], in the NUNM general formalism, is shown on the right plot of figure~\ref{fig:comparison_cc_nc} and compared with the SuperKamiokande analysis. The limit on the $\alpha_{33}$ parameter is derived from~\cite{blennow}; each coloured band in the plot represents the 95$\%$ allowed region. The $\alpha_{33}$ limit found in this analysis is the most precise obtained to date. 
With increasing statistics being collected in larger detector geometries towards the nominal one, further improvements are expected. 

%% file: tex/acknowledgements.tex
\section{Acknowledgements}

The authors acknowledge the financial support of:
KM3NeT-INFRADEV2 project, funded by the European Union Horizon Europe Research and Innovation Programme under grant agreement No 101079679;
Funds for Scientific Research (FRS-FNRS), Francqui foundation, BAEF foundation.
Czech Science Foundation (GAČR 24-12702S);
Agence Nationale de la Recherche (contract ANR-15-CE31-0020), Centre National de la Recherche Scientifique (CNRS), Commission Europ\'eenne (FEDER fund and Marie Curie Program), LabEx UnivEarthS (ANR-10-LABX-0023 and ANR-18-IDEX-0001), Paris \^Ile-de-France Region, Normandy Region (Alpha, Blue-waves and Neptune), France,
The Provence-Alpes-Côte d'Azur Delegation for Research and Innovation (DRARI), the Provence-Alpes-Côte d'Azur region, the Bouches-du-Rhône Departmental Council, the Metropolis of Aix-Marseille Provence and the City of Marseille through the CPER 2021-2027 NEUMED project,
The CNRS Institut National de Physique Nucléaire et de Physique des Particules (IN2P3);
Shota Rustaveli National Science Foundation of Georgia (SRNSFG, FR-22-13708), Georgia;
This work is part of the MuSES project which has received funding from the European Research Council (ERC) under the European Union’s Horizon 2020 Research and Innovation Programme (grant agreement No 101142396).
This work was supported by the European Research Council, ERC Starting grant \emph{MessMapp}, under contract no. $949555$.
The General Secretariat of Research and Innovation (GSRI), Greece;
Istituto Nazionale di Fisica Nucleare (INFN) and Ministero dell’Universit{\`a} e della Ricerca (MUR), through PRIN 2022 program (Grant PANTHEON 2022E2J4RK, Next Generation EU) and PON R\&I program (Avviso n. 424 del 28 febbraio 2018, Progetto PACK-PIR01 00021), Italy; IDMAR project Po-Fesr Sicilian Region az. 1.5.1; A. De Benedittis, W. Idrissi Ibnsalih, M. Bendahman, A. Nayerhoda, G. Papalashvili, I. C. Rea, A. Simonelli have been supported by the Italian Ministero dell'Universit{\`a} e della Ricerca (MUR), Progetto CIR01 00021 (Avviso n. 2595 del 24 dicembre 2019); KM3NeT4RR MUR Project National Recovery and Resilience Plan (NRRP), Mission 4 Component 2 Investment 3.1, Funded by the European Union – NextGenerationEU,CUP I57G21000040001, Concession Decree MUR No. n. Prot. 123 del 21/06/2022;
Ministry of Higher Education, Scientific Research and Innovation, Morocco, and the Arab Fund for Economic and Social Development, Kuwait;
Nederlandse organisatie voor Wetenschappelijk Onderzoek (NWO), the Netherlands;
The grant “AstroCeNT: Particle Astrophysics Science and Technology Centre”, carried out within the International Research Agendas programme of the Foundation for Polish Science financed by the European Union under the European Regional Development Fund; The program: “Excellence initiative-research university” for the AGH University in Krakow; The ARTIQ project: UMO-2021/01/2/ST6/00004 and ARTIQ/0004/2021;
Ministry of Research, Innovation and Digitalisation, Romania;
Slovak Research and Development Agency under Contract No. APVV-22-0413; Ministry of Education, Research, Development and Youth of the Slovak Republic;
MCIN for PID2021-124591NB-C41, -C42, -C43 and PDC2023-145913-I00 funded by MCIN/AEI/10.13039/501100011033 and by “ERDF A way of making Europe”, for ASFAE/2022/014 and ASFAE/2022 /023 with funding from the EU NextGenerationEU (PRTR-C17.I01) and Generalitat Valenciana, for Grant AST22\_6.2 with funding from Consejer\'{\i}a de Universidad, Investigaci\'on e Innovaci\'on and Gobierno de Espa\~na and European Union - NextGenerationEU, for CSIC-INFRA23013 and for CNS2023-144099, Generalitat Valenciana for CIDEGENT/2018/034, /2019/043, /2020/049, /2021/23, for CIDEIG/2023/20, for CIPROM/2023/51 and for GRISOLIAP/2021/192 and EU for MSC/101025085, Spain;
Khalifa University internal grants (ESIG-2023-008, RIG-2023-070 and RIG-2024-047), United Arab Emirates;
The European Union's Horizon 2020 Research and Innovation Programme (ChETEC-INFRA - Project no. 101008324).